\documentclass[12pt]{article}
\usepackage[left=1in,right =1in, top=1in, bottom=1in]{geometry}
\usepackage[linktocpage]{hyperref}
\usepackage[noadjust]{cite}
\usepackage{float}
\usepackage{amsfonts}
\usepackage{caption}
\usepackage{comment}
\usepackage{csquotes}
\usepackage{subcaption}
\usepackage{amsmath}
\usepackage{amssymb}
\usepackage{array}
\usepackage{booktabs}
\usepackage{latexsym}
\usepackage{graphicx}
\usepackage[english]{babel}
\usepackage[font={small}]{caption}
\usepackage{extarrows}
\usepackage{fancybox}
\usepackage{mathtools}
\usepackage{tikz}
\usepackage{adjustbox}

\usepackage[none]{hyphenat}
\usepackage{physics}

\usepackage{url}
\usepackage{enumerate}
\usepackage[shortlabels]{enumitem}
\usepackage{ulem}
\usepackage{tikz}

\begin{document}
\input epsf

\def\p{\partial}
\def\h{{\frac12}}
\def\be{\begin{equation}}
\def\bea{\begin{eqnarray}}
\def\ee{\end{equation}}
\def\eea{\end{eqnarray}}
\def\d{\partial}
\def\la{\lambda}
\def\eps{\epsilon}
\def\b{\bigskip}
\def\m{\medskip}

\newcommand{\newsection}[1]{\section{#1}\setcounter{equation}{0}}

\def\q{\quad}
\def\t{\tilde}
\def\r{\rightarrow}
\def\nn{\nonumber\\}
\def\Ad{\dot{A}}
\def\Bd{\dot{B}}
\def\Cd{\dot{C}}
\def\Dd{\dot{D}}
\def\wb{\bar{w}}
\def\wt{\tilde{w}}
\def\zt{\tilde{z}}
\def\tt{\tilde{t}}

\newcommand{\Romk}[1]{|0^-\rangle_R^{(#1)}}
\newcommand{\Romb}[1]{{}^{\,(#1)}_{\ \,R}\langle0^-|}
\newcommand{\Ropk}[1]{|0^+\rangle_R^{(#1)}}
\newcommand{\Ropb}[1]{{}^{\,(#1)}_{\ \,R}\langle0^+|}
\newcommand{\Rok}[1]{|0\rangle_R^{(#1)}}
\newcommand{\Rob}[1]{{}^{\,(#1)}_{\ \,R}\langle0|}
\newcommand{\Rotk}[1]{|\tilde{0}\rangle_R^{(#1)}}
\newcommand{\Rotb}[1]{{}^{\,(#1)}_{\ \,R}\langle\tilde{0}|}

\newcommand{\Robmk}[1]{|\bar0^-\rangle_R^{(#1)}}
\newcommand{\Robmb}[1]{{}^{\,(#1)}_{\ \,R}\langle\bar0^-|}
\newcommand{\Robpk}[1]{|\bar0^+\rangle_R^{(#1)}}
\newcommand{\Robpb}[1]{{}^{\,(#1)}_{\ \,R}\langle\bar0^+|}
\newcommand{\Robk}[1]{|\bar0\rangle_R^{(#1)}}
\newcommand{\Robb}[1]{{}^{\,(#1)}_{\ \,R}\langle\bar0|}
\newcommand{\Rotbk}[1]{|\bar\tilde{0}\rangle_R^{(#1)}}
\newcommand{\Rotbb}[1]{{}^{\,(#1)}_{\ \,R}\langle\bar\tilde{0}|}

\newcommand{\Roobmk}[1]{|0^-,\bar0^-\rangle_R^{(#1)}}
\newcommand{\Roobmb}[1]{{}^{\,(#1)}_{\ \,R}\langle0^,\bar0^-|}

\def\NSo{0_{\scriptscriptstyle{N\!S}}}
\newcommand{\NSok}[1]{|0\rangle_{\scriptscriptstyle{N\!S}}^{(#1)}}
\newcommand{\NSob}[1]{{}^{\ \,(#1)}_{\ \;\scriptscriptstyle{N\!S}}\langle0|}

\def\phiR{\phi_R}
\def\phiNS{\phi_{\scriptscriptstyle{N\!S}}}
\newcommand{\phiRk}[1]{|\phi\rangle_R^{(#1)}}
\newcommand{\phiRb}[1]{{}^{\,(#1)}_{\ \,R}\langle\phi|}
\newcommand{\phiNSk}[1]{|\phi\rangle_{\scriptscriptstyle{N\!S}}^{(#1)}}
\newcommand{\phiNSb}[1]{{}^{\ \,(#1)}_{\ \;\scriptscriptstyle{N\!S}}\langle\phi|}

\newcommand{\MyRed}{\color [rgb]{0.8,0,0}}
\newcommand{\MyGreen}{\color [rgb]{0,0.7,0}}
\newcommand{\MyBlue}{\color [rgb]{0,0,0.8}}
\newcommand{\MyBrown}{\color [rgb]{0.8,0.4,0.1}}
\newcommand{\MyPurple}{\color [rgb]{0.6,0.0,0.6}}
\def\MH#1{{\MyRed [MH: #1]}}
\def\SM#1{{\MyPurple [SM: #1]}}   
\def\BG#1{{\MyBlue [BG: #1]}}
\def\MM#1{{\MyGreen [MM: #1]}}

\newcommand\blfootnote[1]{%
  \begingroup
  \renewcommand\thefootnote{}\footnote{#1}%
  \addtocounter{footnote}{-1}%
  \endgroup
}

\begin{flushright}
\end{flushright}
\hfill
\vspace{18pt}
\begin{center}
{\Large 
\textbf{Lifting of two-mode states in the D1-D5 CFT
}}

\end{center}

\vspace{10mm}
\begin{center}
{\textsl{Marcel R. R. Hughes$^{a}$}\blfootnote{${}^{a}$hughes.2059@osu.edu}\textsl{, Samir D. Mathur${}^{b}$}\blfootnote{${}^b$mathur.16@osu.edu}\textsl{ and Madhur Mehta${}^{c}$}\blfootnote{${}^c$mehta.493@osu.edu}
\\}

\vspace{10mm}

\textit{\small ${}^{a,b,c}$ Department of Physics, The Ohio State University,\\ Columbus,
OH 43210, USA} \\  \vspace{6pt}

\end{center}

\vspace{12pt}
\begin{center}
\textbf{Abstract}
\end{center}

\vspace{4pt} {\small
\noindent
We consider D1-D5-P states in the untwisted sector of the D1-D5 orbifold CFT where one copy of the seed CFT has been excited by a pair of oscillators, each being either bosonic or fermionic. While such states are BPS at the orbifold point, they will in general `lift' as the theory is deformed towards general values of the couplings. We compute the expectation value of this lift at second order in the deformation parameter for the above mentioned states. We write this lift in terms of a fixed number of nested contour integrals on a given integrand; this integrand depends on the mode numbers of the oscillators in the state. We evaluate these integrals to obtain the explicit value of the lift for various subfamilies of states. At large mode numbers one observes a smooth increase of the lift with the dimension of the state $h$; this increase appears to follow a $\sim \sqrt{h}$ behavior similar to that found analytically in earlier computations for other classes of states.

\vspace{1cm}

\thispagestyle{empty}

\setcounter{footnote}{0}
\setcounter{page}{0}
\newpage

\tableofcontents

\setcounter{page}{1}

\numberwithin{equation}{section}

\newpage
\section{Introduction}\label{secintro}
\normalem

In string theory D1-branes, D5-branes and momentum (P) charges form an interesting bound state which has been extensively studied, particularly in the context of understanding black hole microstates. A solution of classical gravity with the same charges and mass exists; this hole has a Bekenstein entropy of $S_{\rm{bek}}=A/4G$. In this case -- a brane system preserving a fraction of the supersymmetries of string theory -- the microstates of the hole will be BPS bound states with these charges. A lower bound on the number of these BPS states can be obtained using an index which was computed in \cite{Strominger:1996sh} for type IIB string theory compactified on $K3\times S^1$ and in \cite{Maldacena:1999bp} for IIB compactified on $T^4\times S^1$. In each of these two cases the leading-order behavior of the index matches the Bekenstein entropy.

The construction of these black hole microstates in the gravitational description is termed the fuzzball program. A large set of microstates have been constructed (see \cite{Bena:2022ldq,Bena:2022rna} for reviews of the current state of the fuzzball program, as well as the related microstate geometry program). This program of work has also been able to provide a holographic dictionary between these gravity states and those of the dual D1-D5 CFT  \cite{Lunin:2001jy,Mathur:2005zp,Kanitscheider:2007wq,Bena:2007kg,Chowdhury:2010ct,Shigemori:2020yuo}. It is interesting to examine the states in the CFT description to understand which states are BPS at a general point in moduli space and therefore  correspond to microstates of the extremal black hole. 

Considering these black hole microstates at a deeper level reveals some interesting questions. The conformal field theory (CFT) dual to this brane system -- the so-called D1-D5 CFT -- is conjectured to have an `orbifold point' in its  moduli space where it has a description in terms of a ($1+1$)-dimensional sigma model with a symmetric group orbifold target space \cite{Vafa:1995bm,Dijkgraaf:1998gf,Larsen:1999uk,Seiberg:1999xz,Arutyunov:1997gi,Arutyunov:1997gt,Jevicki:1998bm,David:2002wn}, analogous to free super Yang-Mills (SYM) in the D3-brane system \cite{Maldacena:1997re}. At this orbifold point any states of the theory with purely left-moving (or purely right-moving) excitations is a BPS state -- some fraction of the total $\mathcal{N}=4$ supersymmetry of the CFT is preserved. Deforming the D1-D5 CFT away from the orbifold point and towards the strongly-coupled part of the moduli space, for which there is a description in terms of a dual semiclassical gravity theory, some of the short multiplets formed from these BPS states join together into long multiplets and `lift'. This lift is  an anomalous contribution to the conformal dimension of these states. Clearly only states that remain unlifted (remain BPS) when moving away from the orbifold point can be BPS at general values of the moduli and can thus contribute to the index count of states. Given that the {\it number} of states that must remain  unlifted is given by an index \cite{Strominger:1996sh,Maldacena:1999bp} an important question to ask is: {\it which} states are unlifted and why? 

A puzzling phenomenon can be stated in the computation of the lifts of CFT states. In \cite{Gava:2002xb}, Gava and Narain found that almost all low-energy states of the CFT were lifted; this is in line with expectations from the gravity theory where the only low-lying BPS states describe supergravity quanta, while the rest describe string states whose energies are lifted up to the string scale. However, we know that we will reach a black hole phase in the CFT spectrum at high enough energies; in this domain there must be  a large number of unlifted states in order to account for the large index calculation results. The question is then: what changes when we look at states above the black hole threshold in these lifting computations? Answering this question would also tell us about the nature and properties of microstates that contribute to the extremal black hole's entropy. No simple answer to this question has yet been found\footnote{Some recent progress has been made in the case of microstates of AdS${}_5$ black holes from the study of states in $\mathcal{N}=4$ SYM with gauge group $SU(2)$ above the black hole threshold~\cite{Chang:2022mjp}.}. The progress made so far has consisted of computing the lift of various families of D1-D5-P states. Some patterns have emerged from these computations and it is hoped that a general answer to the above question will become clear if we understand the lift in enough special cases. The families for which the lift has been computed so far include the following\footnote{See \cite{Lima:2020boh,Lima:2020kek,Lima:2020nnx,Lima:2020urq,Lima:2021wrz,Lima:2021xqj,Lima:2022cnq,Benjamin:2022jin,Guo:2022and} for additional work on lifting of states and conformal perturbation theory in the D1-D5 CFT and \cite{Guo:2022sos,Guo:2023czj} for work related to twist operators.}:
\begin{enumerate}

\item \label{lift0} In \cite{Gava:2002xb} a leading order lift was computed for general states of the following kind. We start with all copies of the $c=6$ CFT singly wound, and in their NS ground state; this is dual to the gravity solution $AdS_3\times S^3\times T^4$. Then we take $k$ copies of the $c=6$ CFT with $k\gg 1$ and twist them together to make a multi-wound component string. This component string is then excited by left moving excitations to a level $n\ll k$. All such states were found to be lifted, apart from those that correspond to the single particle supergravity multiplet.

\item \label{lift1} In \cite{Guo:2022ifr} the untwisted-sector states considered were built from the global NS-NS vacuum with one copy excited with a primary of the single-copy (small) $\mathcal{N}=4$ superconformal algebra. This family of states is labelled by the holomorphic dimension $h$ and the second-order lift was found to be
\begin{equation}
    E^{(2)}_{h} = (N-1)\frac{\lambda^2\pi^2}{2^{2h-1}} \frac{\Gamma(2h)}{\Gamma(h)^2} \ .
\end{equation}
It was shown that explicit knowledge of the form of the superconformal primary is not necessary, only its conformal dimension.

\item \label{lift2} In \cite{Guo:2019ady,Guo:2020gxm}, for the case of $N=2$, all D1-D5-P primaries up to level 4 were explicitly constructed and their lifts computed on a case-by-case basis. It was found that all such states that were able to lift, did do so\footnote{States that are BPS at the orbifold point of the theory are in short multiplets of the symmetry algebra. In order for such states to lift, four of these short multiplets must join into one long multiplet of the deformed theory. A state being allowed to lift means that there exist suitable short multiplets that can join.}. The observation was made that of the states that were not lifted, a large proportion were from the doubly-twisted sector.

\item \label{lift3} In \cite{Hampton:2018ygz} the specific family of states formed by exciting the global NS-NS vacuum with the excitation $J^+_{-(2m-1)}\cdots J^+_{-3}J^+_{-1}$ on $n$ out of the $N$ copies was considered. These states have dimensions $(h,\bar{h})=(n m^2,0)$ and $j^3_0$ charges $(j,\bar{j})=(n m,0)$ and their second-order lift was found to be
\begin{equation}
    E^{(2)}_{m,n} = n(N-n)\frac{\lambda^2 \pi^{\frac32}}{2} \frac{\Gamma\big(m^2-\frac12\big)}{\Gamma(m^2-1)} \ .
\end{equation}

\item \label{lift4} In \cite{Gaberdiel:2015uca} the currents of the higher-spin symmetry algebra found at the orbifold point were studied. This enhanced symmetry is broken to the usual $\mathcal{N}=4$ superconformal algebra when moduli associated with the string tension are turned on. These currents are then lifted and the pattern of their anomalous dimensions was studied and Regge trajectories were identified.

\item \label{lift5} In \cite{Benjamin:2021zkn} the lift of untwisted-sector $1/4$-BPS states in $(h,j)=(1,0)$ left-moving long multiplets and general $\bar{j}$ right-moving short multiplets was systematically computed for general $N$. Further evidence was found that the supersymmetric index computations undercount the number of unlifted $1/4$-BPS states, particularly at large $N$.

\end{enumerate}

In the present paper, we will compute the lift for a new set of states, described as follows. We start with the D1-D5 theory in the untwisted sector, with each copy of the CFT in the NS-NS vacuum state; the dual gravity state for this is $AdS_3\times S^3\times T^4$. We now excite {\it one} of the copies of the CFT (still keeping it untwisted). The excitation is given by the application of a pair of left-moving oscillators; either the purely bosonic excitations $ \alpha_{A\dot A, -m} \alpha_{B\dot B,-n}$, the mixed case $ \alpha_{A\dot A, -m} d^{\beta B}_{-n}$ or the purely fermionic excitations $ d^{\alpha A}_{-m} d^{\beta B}_{-n}$. We compute the expectation value of the energy $\langle E\rangle$ for such states to order $\lambda^2$ in the deformation away from the orbifold point (the difference between this and the energy at the orbifold point is termed the lift). 

Note that these state are not in general primaries of the Virasoro algebra for one copy of the CFT; thus they differ from the class (\ref{lift2}) described above. We will also look at energy levels, enumerated by the mode numbers $m$ and $n$, that are significantly higher than the energy levels studied in case (\ref{lift2}), so we hope to capture the pattern of the lift of such states for general energies $m,n$.

A brief summary of our findings is as follows: in Section~\ref{sec.main} we derive the lifts for the above mentioned families of states.  We give an expression for the lift for arbitrary states of this type, in terms of  a fixed number of nested contour integrals on a given integrand; this integrand depends on the mode numbers of the oscillators in the state. We evaluate these integrals to obtain the explicit value of the lift for various subfamilies of states.  These lifting computations hint at a $\sqrt{h}$ behavior for large $h$, where $h$ is the dimension of the state under consideration. This $\sqrt{h}$ behavior has previously been seen analytically in \cite{Hampton:2018ygz,Guo:2022ifr}. However, due to the complexity of the current computation, it is difficult to find a closed form expression for the lifts considered. In Section~\ref{sec.checks} we derive relations between lifts stemming from Ward identities of the superconformal algebra on the lifting calculations. It is shown how these various relations are satisfied by the explicit lifts computed in Section~\ref{sec.main}, serving as non-trivial checks of the method.

\section{The D1-D5 CFT and its free-orbifold point} \label{secD1D5}

We consider type IIB string theory with a background compactified as
\begin{equation} \label{eq.background}
    M_{9,1}\r M_{4,1}\times S^1\times T^4 \ ,
\end{equation}
with $n_1$ D1-branes and $n_5$ D5-branes wrapped on the $S^1$ and with the D5-branes wrapped also on the $T^4$. The bound states of this brane system in the IR generate the D1-D5 CFT, which is a $(1+1)$-dimensional conformal field theory living on a cylinder made from the time direction and the spatial $S^1$. This theory is believed to have a point (more correctly, a locus) in its moduli space, called the free-orbifold point, where we have a description of the theory in terms of
\be
    N = n_1 n_5 \ ,
\ee
copies of a free seed $c=6$ CFT. This free CFT contains $4$ free bosons $\{\partial X\}$ and $4$ free fermions in the left-moving sector $\{\psi\}$ and likewise for the right-moving sector $\{\bar{\psi}\}$. These free fields are subject to an orbifolding by the group of permutations $S_N$, leading to the Hilbert space factoring into twisted sectors labelled by $1\leq k\leq N$. These different sectors effectively each describe a CFT on a $k$-wound circle -- sometimes referred to as a component string. This orbifold point of the D1-D5 CFT has been shown~\cite{Giribet:2018ada,Gaberdiel:2018rqv,Eberhardt:2018ouy,Eberhardt:2019ywk,Eberhardt:2020akk,Eberhardt:2021vsx} to be dual to a string in an AdS${}_3\times S^3\times T^4$ background in the tensionless limit with one unit of NS-NS flux. Extensions of these ideas can be found, for example, in~\cite{Eberhardt:2019qcl,Dei:2019osr,Brollo:2023pkl,Brollo:2023rgp}. It is thought that this free-orbifold theory can be deformed in the moduli space of the D1-D5 CFT towards a strong coupling regime, at which the theory would have a dual semiclassical gravity description.

\subsection{Symmetries of the CFT} \label{ssec.symmCFT}

The D1-D5 CFT has $\mathcal{N}=4$ supersymmetry in both the left- and right-moving sectors~\cite{Schwimmer:1986mf,Sevrin:1988ew} at a generic point in its moduli space. This leads to an ${\cal N}=4$ superconformal symmetry algebra in both the left and right sectors, with chiral algebra generators
\begin{equation} \label{eq.lcurr}
    L_{n}\ \ ,\quad G^{\alpha}_{\dot A,r}\ \ ,\quad J^a_n \ ,
\end{equation}
for the left movers associated with the stress-energy tensor, supercurrents and $\mathfrak{su}(2)$ R-currents respectively. The right-moving sector has analogous generators given by
\begin{equation} \label{eq.rcurr}
    \bar L_{n}\ \ ,\quad \bar G^{\bar \alpha}_{\dot A,r} \ \ ,\quad \bar J^a_n \ .
\end{equation}
The indices $\alpha$ and $\bar{\alpha}$ are doublet indices of the $SU(2)_L$ and $SU(2)_R$ factors of the $SO(4)_E\cong (SU(2)_L\times SU(2)_R)/\mathbb{Z}_2$ R-symmetry of the $\mathcal N=4$ superconformal algebra, where the subscript $E$ stands for `external', which denotes the geometric origin of this symmetry from rotations in the noncompact spatial directions of the background \eqref{eq.background}. There is also another $SO(4)$ global symmetry coming from the $T^4$ factor of the background which we call the `internal' $SO(4)_I$. This symmetry is broken by the compactification on the torus, but at the orbifold point it still provides a useful organising principle from the spectrum. We write $SO(4)_I\cong (SU(2)_1\times SU(2)_2)/\mathbb{Z}_2$ and use doublet indices $A, \dot A$ for $SU(2)_1$ and $SU(2)_2$ respectively. In \eqref{eq.lcurr} and \eqref{eq.rcurr} the index $a$ is a vector index of $SO(4)_E$. This symmetry algebra is in fact enlarged to the contracted large ${\cal N}=4$ superconformal symmetry \cite{Maldacena:1999bp,Sevrin:1988ew} which includes the free bosons and fermions of the orbifold theory that we write as
\begin{equation} \label{eq.freeFields}
    \partial X_{A\Ad} \quad,\quad \psi^{\alpha A}\quad,\quad \bar{\psi}^{\bar{\alpha}\Ad} \ .
\end{equation}
This large $\mathcal N=4$ superconformal algebra and our conventions are outlined in Appendix~\ref{app_cft}. Exactly at the free point, this symmetry is in fact boosted to include also a $\mathcal{W}_{\infty}$ algebra studied, for instance, in~\cite{Gaberdiel:2015mra,Gaberdiel:2015uca}.

\subsection{Neveu-Schwarz and Ramond sectors} \label{sec.sf}

In two dimensions the $\mathcal{N}=4$ superconformal algebra additionally has a global automorphism group, part of which is referred to as spectral flow transformations\footnote{The full automorphism group of this $\mathcal{N}=4$ superconformal algebra is in fact $SO(4)$ with spectral flows being a one-parameter subgroup~\cite{Schwimmer:1986mf}.}. These spectral flow transformations map between equivalent algebras in which the periodicity of the various symmetry generators are changed. Here we only very briefly describe some aspects of spectral flow transformations since we will require very little of this machinery in this paper.

Under a spectral flow transformation by $\eta$ units the conformal dimension $h$ and $J^3_0$ charge $m$ of an operator transform as%
\footnote{It should be noted that the left- and right-moving sectors of the $\mathcal{N}=(4,4)$ superconformal algebra can be spectrally flowed by $\eta$ and $\bar{\eta}$ units independently. In~\cite{Guo:2021uiu} a generalisation was studied in which the generators of the free-field realisation of the $\mathcal{N}=4$ algebra are decomposed into generators of two $\mathcal{N}=2$ algebras, which one can spectral flow separately. This was termed `partial spectral flow' but we will not have use for it here.}%
\begin{equation} \label{sfDims}
    h \rightarrow h' = h +m\eta +\frac{c \eta^2}{24} \quad\ ,\qquad m\rightarrow m' = m + \frac{c\eta}{12}\ ,
\end{equation}
where $c$ is the central charge of the CFT. Strictly speaking, the value of $c$ used in the above transformation rules depends on the number of copies on which the state being transformed acts and how these copies are twisted together. For instance, if the state is on a single copy then $c=6$. In general, a state on a component string of $k$ copies twisted together the central charge would be $c=6k$.

In appendix C of \cite{Guo:2022ifr} it was shown that a superconformal primary field $\phi(z)$ of the $\mathcal{N}=4$ algebra%
\footnote{An NS-sector superconformal primary is defined by being annihilated by all positive modes of the current algebra
\begin{equation} \label{def primary 2}
    L_{n}|\phi\rangle_{\scriptscriptstyle{N\!S}} = G^{\alpha}_{\dot A,r}|\phi\rangle_{\scriptscriptstyle{N\!S}} = J^{a}_{n}|\phi\rangle_{\scriptscriptstyle{N\!S}} =0 \quad,\quad n>0\ ,\ r\geq \frac12 \ .
\end{equation}
}%
on the plane transforms under spectral flow as
\begin{equation} \label{sf_01}
    \phi(z)\to (z-z_0)^{-\eta q}\phi(z)\ ,
\end{equation}
with $q$ being the primary's $j^3_0$ charge. For an odd integer value of $\eta$, spectral flow switches between the NS- and R-sector fermionic boundary conditions for the theory. Importantly for the present paper, the basic bosons and left-moving fermions of the free orbifold theory \eqref{eq.freeFields} transform in the following manner under spectral flow by $\eta$ units around $z_0$ on the plane
\begin{equation} \label{freeFieldSF}
     \partial X_{A\Ad}(z)\to\partial X_{A\Ad}(z) \quad,\quad \psi^{\alpha A}(z)\to (z-z_0)^{-\eta q_\alpha} \psi^{\alpha A}(z) \ ,
\end{equation}
where $q_\alpha=\pm\frac12$ is the $J^3_0$ charge of $\psi^{\alpha A}(z)$.

\subsection{Deformation away from the orbifold point}

The D1-D5 CFT can be deformed away from the free orbifold point by the addition of the exactly marginal operator $D$ to the Lagrangian, modifying the action as
\begin{equation} \label{defor S}
    S\r S+\lambda \int d^2 z\, D(z, \bar z) \ ,
\end{equation}
where $D$ has conformal dimensions $h=\bar h=1$. The theory has a total of $20$ exactly marginal operators; $16$ from the shape and complex structure moduli of $T^4$ and $4$ from superdescendants of the twist-2 chiral primaries $\sigma^{\alpha\bar\alpha}$ in the orbifold theory~\cite{Lunin:2001fv}. The $T^4$ moduli are `trivial deformations', whereas the latter four break the additional higher-spin symmetry found only at the orbifold point and are also the direction towards the region with a semi-classical gravity description. The particular deformation operator $D$ of this type that is a singlet under all $SU(2)$ symmetries of the orbifold theory is
\begin{equation} \label{D 1/4}
    D = \frac{1}{4}\epsilon^{\dot A\dot B}\epsilon_{\alpha\beta}\epsilon_{\bar\alpha \bar\beta} \,G^{\alpha}_{\dot A, -\h} \bar G^{\bar \alpha}_{\dot B, -\h} \sigma^{\beta \bar\beta} \ ,
\end{equation}
where $G$ and $\bar G$ are left- and right-moving supercharge modes at the \emph{orbifold point}, \textit{i.e.} at $\lambda=0$. The remaining three non-trivial deformation operators are in the triplet projection of $SU(2)_2$.

\subsection{Lift formulae} \label{sec:LiftForm}

The method of computation of lifts used in this paper was developed in~\cite{Hampton:2018ygz} by the use of conformal perturbation theory. Here we do not give details of the derivation of this method and instead use it as a tool. This method requires the computation of the second order lift from the integrated correlator
\begin{equation}
    A^{(2)}(T) = \frac12 \Big<\phi\big(\tfrac{T}{2}\big)\Big| \int d^2w_2 D(w_2,\bar{w}_2) \int d^2w_1 D(w_1,\bar{w}_1) \Big|\phi\big(\!-\!\tfrac{T}{2}\big)\Big> \ ,
\end{equation}
where the factor of $\tfrac12$ comes from the second order perturbation in the path integral and the initial and final states are placed at finite $\tau$ in order to regularise the calculation. It is convenient to write these two insertions of the deformation operator $D(w,\wb)$ in different forms
\begin{equation} \label{eq.deformationDef}
    D(w,\bar{w}) = \epsilon^{\dot A \dot B} G^-_{\dot A,-\frac12} \bar{G}^-_{\dot B,-\frac12} \sigma^{++}_2(w,\bar w) = \epsilon^{\dot A \dot B} G^+_{\dot A,-\frac12} \bar{G}^+_{\dot B,-\frac12} \sigma^{--}_2(w,\bar w) \ ,
\end{equation}
which are equivalent to \eqref{D 1/4} but with the explicit $SU(2)$ singlet structure obscured.
\begin{figure}[h]
    \centering
    \includegraphics[scale =0.55]{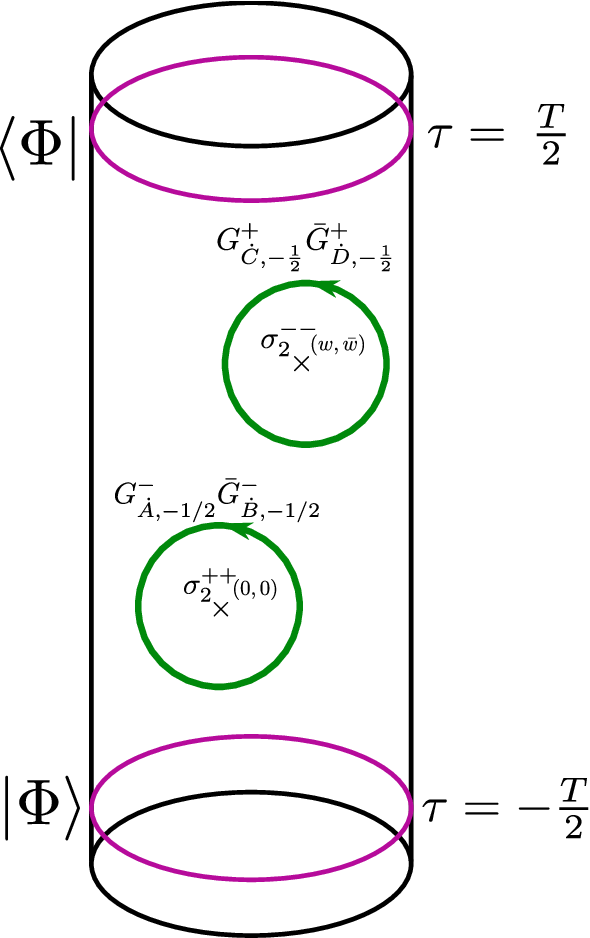}
    \caption{In the above figure, we depict the correlator from \eqref{eq.Xdef}. In purple, we see the initial $\ket{\Phi}$ and final states $\bra{\Phi}$ located at $\tau =-T/2$ and $\tau = T/2$ respectively. The green contours are the supersymmetry modes from the two deformation operators placed at $(0,0)$ and $(w,\bar{w})$.}
    \label{fig:cylinderstate}
\end{figure}
Fixing the insertion of the first deformation operator to the origin and considering the large $T$ limit we can write
\begin{equation}\label{amplitude}
    A^{(2)}(T) \approx \frac12 (2\pi T) e^{-E^{(0)} T} \big<\Phi\big| \int d^2w_2 D(w_2,\bar{w}_2) D(0,0) \big|\Phi\big> \ .
\end{equation}
From conformal perturbation theory it was shown that the second order lift in energy is given by the coefficient of $-Te^{-E^{(0)}T}$ in the correlator $A^{(2)}(T\to\infty)$ and so
\begin{equation} \label{eq.lift}
    E^{(2)}\big(|\Phi\rangle\big) = -\pi \lambda^2 \lim_{T\to\infty} \frac{X(T)}{\langle\Phi|\Phi\rangle} \ ,
\end{equation}
where we define
\begin{equation}\label{eq.Xdef}
    X(T) \equiv \epsilon^{\dot C\dot D}\epsilon^{\dot A \dot B} \int d^2w\, \big<\Phi\big| \big(G^+_{\dot C,-\frac12}\bar{G}^+_{\dot D, -\frac12}\sigma^{--}_2\big)(w,\bar w) \big(G^-_{\dot A,-\frac12}\bar{G}^-_{\dot B, -\frac12}\sigma^{++}_2\big)(0,0)  \big|\Phi\big> \ ,
\end{equation}
with the initial and final states are located at $\tau =-\frac{T}{2}$ and $\tau = \frac{T}{2}$ respectively. The configuration of the amplitude in \eqref{eq.Xdef} is depicted in Figure~\ref{fig:cylinderstate}. In \eqref{eq.Xdef} we have used two different forms of the deformation operators for the two insertions, as given in \eqref{eq.deformationDef}. Since the right-moving part of the states we consider is just the NS vacuum on all copies, the right-moving part of the correlator in \eqref{eq.Xdef} will be determined simply by the $D\times D$ OPE and thus will contribute a factor independent of the initial and final states of the form
\begin{equation} \label{eq.RmovingCorr}
    \langle\NSo|\big(\bar{G}^{+}_{\Dd,-\frac12}\sigma^-_2\big)(\wb) \big(\bar{G}^{-}_{\Bd,-\frac12}\sigma^{+}_2\big)(0) |\NSo\rangle  = \frac{-\epsilon_{\dot D\dot B}}{4\sinh^2(\frac{\bar w}{2})} \ .
\end{equation}
It is convenient to then write $X(T)$ as
\begin{align} \label{eq.XT}
    X(T) &= \epsilon^{\dot C\dot D}\epsilon^{\dot A \dot B} \int d^2w\, A(w,0)\epsilon_{\Ad\Cd}  \bigg(\frac{-\epsilon_{\dot D\dot B}}{4\sinh^2(\frac{\bar w}{2})}\bigg) \nonumber\\
    &= -\frac12\epsilon^{\dot C\dot D}\epsilon^{\dot A \dot B}\epsilon_{\dot D\dot B}\epsilon_{\Cd\Ad} \int d^2w\, A(w,0)\, \partial_{\wb}\big(\coth(\tfrac{\bar w}{2})\big) \nonumber\\
    &= \frac{i}{2} \int_C dw\, A(w,0) \coth(\tfrac{\bar w}{2}) \nonumber\\
    &\equiv I_{C_1} + I_{C_2} + I_{C_3} \ ,
\end{align}
where the contours $C_1$, $C_2$ and $C_3$ on the cylinder (depicted in Figure~\ref{fig:ICs}) are at $\tau=\frac{T}{2},-\frac{T}{2}$ and around $|w|=\epsilon$ respectively and $A(w_2,w_1)$ is simply the non-integrated left-moving amplitude
\begin{equation}
    A(w_2,w_1) \equiv \big<\Phi\big| \Big(G^+_{-,-\frac12}\sigma^-_2\Big)(w_2)\Big(G^-_{+,-\frac12}\sigma^+_2\Big)(w_1)\big|\Phi\big> \ ,
\end{equation}
where specific $SU(2)_2$ indices on the $G$ modes have been chosen for ease of computation.
\begin{figure}[H]
    \centering
    \includegraphics[scale =0.53]{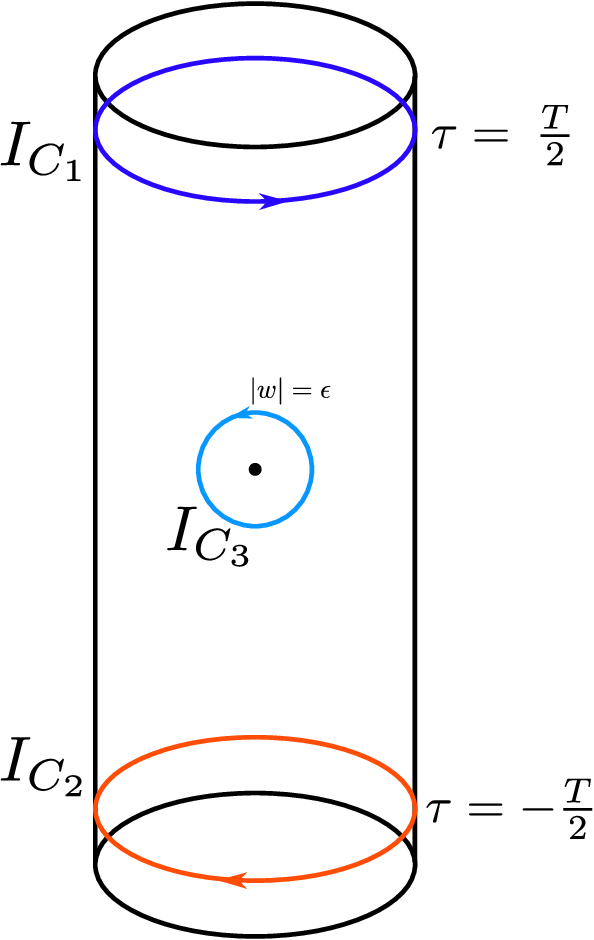}
    \caption{We show the three boundary contour integrals $I_{C_1}, I_{C_2}$ and $I_{C_3}$ defined in equations \eqref{eq.IC1}, \eqref{eq.IC2} and \eqref{eq.IC3} respectively. Together these contour integrals yield the integrated amplitude in \eqref{eq.XT} required in the lift \eqref{eq.lift}.} \label{fig:ICs}
\end{figure}
The three contour integrals in \eqref{eq.XT} are then
\begin{subequations} \label{eq.IC123defs}
    \begin{align}
        I_{C_1}(T) &\equiv - \frac12\int^{2\pi}_0 \!d\sigma\, A(w,0) \coth(\tfrac{\bar w}{2}) \ ,\label{eq.IC1} \\
        I_{C_2}(T) &\equiv \frac12\int^{2\pi}_0 \!d\sigma\, A(w,0) \coth(\tfrac{\bar w}{2}) \ ,\label{eq.IC2} \\
        I_{C_3} &\equiv -\frac{i}{2} \oint_{|w|=\epsilon} \!dw\, A(w,0) \coth(\tfrac{\bar w}{2}) \ ,\label{eq.IC3}
    \end{align}
\end{subequations}
where the complex coordinates on the cylinder have been written as $w=\tau+i\sigma$ and $\wb=\tau-i\sigma$. In $I_{C_3}$ the cutoff $\epsilon$ should be taken to $0$ at the end of the calculation. The computation thus boils down to evaluating the non-integrated left-moving amplitude $A(w,0)$ with the universal right-moving $D\times D$ OPE contribution factored out. We note that, as discussed in \cite{Hampton:2018ygz}, the integral $I_{C_3}$ contributes only a universal divergent piece $\sim \frac1{\epsilon}$ independently of the choice of external states which is removed by a counterterm in the action. Thus the integral $I_{C_3}$ does not contribute to the lift \eqref{eq.lift}.

\section{Lifting of single-copy two-mode states} \label{sec.main}

\subsection{The states} \label{sec.states}

In this paper we consider untwisted sector states formed by acting with two modes of the free bosons and fermions of the orbifold theory on the NS-NS vacuum on one copy $\NSok{1}$, with the remaining copies being in the ground state. These two-mode states fit into three categories of states
\begin{equation} \label{eq.states}
    \alpha_{B\Bd,-m}\alpha_{A\Ad,-n}|\NSo\rangle \quad,\quad \alpha_{B\Bd,-m} d^{\alpha A}_{-s}|\NSo\rangle \quad,\quad d^{\beta B}_{-r} d^{\alpha A}_{-s}|\NSo\rangle \ ,
\end{equation}
which we shall define precisely in their respective sections. The computation of the lifts of these three groups of states will be treated separately below. These two-mode families of states are the simplest non-trivial choices since for all single-mode states, \textit{i.e.} $\alpha^{(1)}_{A\Ad,-n}|\NSo\rangle$ and $d^{(1)\alpha A}_{-s}|\NSo\rangle$, the lift vanishes. This is because in the process of forming copy-symmetric states from them we obtain a global mode\footnote{A global mode here refers to one that acts diagonally in copies, \textit{i.e.} $\mathcal{O}_n^{global}=\sum_{i=1}^N\mathcal{O}_{n}^{(i)}$ where $\mathcal{O}_{n}^{(i)}$ acts purely on the $i$th copy.} acting on the global NS-NS vacuum, whose lift is zero~\cite{Hampton:2018ygz}. This is not the case for states of the form \eqref{eq.states} or those formed from more than two modes.

Since the computation of the lift sees only a single pair of copies at a time\footnote{This is due to the twist-$2$ fields in the deformation operators acting on a pair of copies and the fact that the lift is an expectation value, forcing the second of the twist operators to simply act on the same pair as the first.} it is sufficient at this stage to consider a specific ordered pair of copies with one copy being excited and one copy in the NS-NS vacuum. As explained in Section (3.1) of \cite{Guo:2022ifr}, the extrapolation of the final lifting results to the case of generic $N$ is then a matter of combinatorics to get
\begin{equation} \label{eq.GeneralN}
    E^{(2)}_N\big(|\Phi\rangle\big) = 2(N-1) E^{(2)}\big(|\Phi\rangle\big) \ ,
\end{equation}
where $E^{(2)}(|\Phi\rangle)$ is the two-copy lift that we work with in this paper.

\subsection{Lifting of $|\alpha\alpha\rangle_{\!B\Bd A\Ad(m,n)}$} \label{ssec.Liftaa}

Firstly we consider the lift of states of the form
\begin{equation} \label{eq.aaState}
    |\alpha\alpha\rangle_{\!B\Bd A\Ad(m,n)} \equiv \frac{1}{\sqrt{2}} \Big(\alpha^{(1)}_{B\Bd,-m}\alpha^{(1)}_{A\Ad,-n} + \alpha^{(2)}_{B\Bd,-m}\alpha^{(2)}_{A\Ad,-n}\Big) \NSok{1}\NSok{2} \ ,
\end{equation}
where the factor of $\tfrac1{\sqrt{2}}$ comes from the normalisation of the state over two copies. The norm of this state can be found from the following inner product
\begin{align} \label{eq.aaNormCorr}
    {}_{(m,n)D\Dd C\Cd}\langle \alpha\alpha|\alpha\alpha\rangle_{\!B\Bd A\Ad(m,n)}
    &= mn\big(\epsilon_{AC}\epsilon_{BD}\epsilon_{\Ad\Cd}\epsilon_{\Bd\Dd}\delta_{m,n} + \epsilon_{AD}\epsilon_{BC}\epsilon_{\Ad\Dd}\epsilon_{\Bd\Cd}\big) H[m-1] \nonumber\\
    &\quad + 2 m^2 \epsilon_{AB}\epsilon_{CD}\epsilon_{\Ad\Bd}\epsilon_{\Cd\Dd} \delta_{m+n,0} H[-m-1] \ ,
\end{align}
under the condition that for the bra state in \eqref{eq.aaNormCorr} to be the Hermitian conjugate of the ket all $SU(2)$ indices should be opposite to those of the ket and additional negative signs are included as per the conjugation conventions given in \eqref{eq.adDagConv}. In \eqref{eq.aaNormCorr} we have used the discrete step function definition
\begin{equation}
    H[n] \equiv \begin{cases}
        \ \ 1 \quad \text{for } n\geq0\\
        \ \ 0 \quad \text{for } n<0
    \end{cases}\ .
\end{equation}
The computation of the lift requires the following left-moving amplitude
\begin{equation} \label{eq.Aaa11def}
    A_{m,n}^{(1)(1)}(w_2,w_1) \equiv \langle\NSo|\big(\alpha^{(1)}_{D\Dd,n}\alpha^{(1)}_{C\Cd,m}\big)\, \Big(G^+_{-,-\frac12}\sigma^-\Big)(w_2) \Big(G^-_{+,-\frac12}\sigma^+\Big)(w_1)\, \big(\alpha^{(1)}_{B\Bd,-m}\alpha^{(1)}_{A\Ad,-n}\big) |\NSo\rangle \, ,
\end{equation}
where we have chosen all of the $\alpha$ modes to act on copy 1 and have suppressed the $SU(2)_2$ indices on the left-hand side for easy of notation. The external states will need to be symmetrised over copy indices later. Clearly in order to have a non-vanishing initial state we require the condition
\begin{equation}
    n>0 \ .
\end{equation}
As described in Section~\ref{sec:LiftForm} the method of computing this amplitude is to map it from the (doubly covered) cylinder to the (doubly covered) plane, from which we map to the covering space\footnote{As found in~\cite{Lunin:2000yv} and used in the context of lifting in~\cite{Hampton:2018ygz} the covering space map $z\to t$ for correlators containing two order-2 twist operators is given by $z(t)=\frac{(t+a)(t+b)}{t}$. The insertions of the two deformation operators on the plane $z_1=e^{w_1}$ and $z_2=e^{w_2}$ are mapped to the points $t_1=-\sqrt{ab}$ and $t_2=\sqrt{ab}$ on the cover. Our conventions are that the points $z=0$ and $z=\infty$ on the first sheet of the doubly-covered plane map to $t=-a$ and $t=\infty$ on the covering space.} in order to geometrically resolve the effect of the two twist operator insertions. Under these two maps the initial state transforms as
\begin{align} \label{eq.aatransi}
    \big(\alpha^{(1)}_{B\dot B,-m} \alpha^{(1)}_{A\dot A,-n}\big)_{-\infty} &= i^2 \oint_{-\infty}\frac{dw_4}{2\pi i} \frac{dw_3}{2\pi i}\, e^{-m w_4}e^{-n w_3} \, \partial X_{B\dot{B}}(w_4) \partial X_{A\dot{A}}(w_3) \nonumber\\
    &= i^2 \oint_{0}\frac{dz_4}{2\pi i} \frac{dz_3}{2\pi i}\, z_4^{-m} z_3^{-n} \, \partial X_{B\dot{B}}(z_4) \partial X_{A\dot{A}}(z_3) \nonumber\\
    &= i^2 \oint_{-a}\frac{dt_4}{2\pi i} \frac{dt_3}{2\pi i}\, z_4(t_4)^{-m} z_3(t_3)^{-n} \, \partial X_{B\dot{B}}(t_4) \partial X_{A\dot{A}}(t_3) \ ,
\end{align}
and likewise the final state transforms as
\begin{align} \label{eq.aatransf}
    \big(\alpha^{(1)}_{D\dot D,n}\alpha^{(1)}_{C\dot C,m}\big)_{\infty} &= i^2 \oint_{\infty}\frac{dw_5}{2\pi i} \frac{dw_6}{2\pi i}\, e^{m w_5}e^{n w_6} \, \partial X_{D\Dd}(w_6) \partial X_{C\Cd}(w_5) \nonumber\\
    &= i^2 \oint_{\infty}\frac{dz_5}{2\pi i} \frac{dz_6}{2\pi i}\, z_5^{m} z_6^{n} \, \partial X_{D\Dd}(z_6) \partial X_{C\Cd}(z_5) \nonumber\\
    &= i^2 \oint_{\infty}\frac{dt_5}{2\pi i} \frac{dt_6}{2\pi i}\, z_5(t_5)^{m} z_6(t_6)^{n} \, \partial X_{D\Dd}(t_6) \partial X_{C\Cd}(t_5) \ .
\end{align}
On mapping to the covering space the order-2 twist fields are resolved, leaving behind spin fields $S^-_2(t_2)$ and $S^-_2(t_1)$ which can in turn be removed by spectrally flowing on the $t$-plane by $\eta=-1$ units around $t=t_1$ and by $\eta=1$ units around $t=t_2$. Since the $\alpha$ modes have no charge under $J^3_0$ these initial and final states transform trivially under these two spectral flows. The transformation of the $\sigma$ fields inserted at $w_1$ and $w_2$ on the cylinder under these maps and spectral flows are universal to all of our lifting calculations and thus are packaged into the base amplitude
\begin{equation} \label{eq.baseAmp}
    U(w_2,w_1) \equiv \langle\NSo| \sigma^-(w_2)\sigma^+(w_1)|\NSo\rangle = \frac{1}{2\sinh \!\big(\frac{w_2-w_1}{2}\big)} \ ,
\end{equation}
which was derived in~\cite{Hampton:2018ygz}, along with factors from the transformation of the $G$ modes\footnote{See Section (4.3) of \cite{Hampton:2018ygz} for a derivation of the transformation factors for the $G$ modes coming from the two deformation operators under the maps to the $t$-plane and spectral flows. In total these factors are $\big[t_1t_2 (t_1+a)(t_2+a)(t_1+b)(t_2+b)\big]^{\frac12}$.}. This leaves the amplitude \eqref{eq.Aaa11def} (up to the overall factors described above and the much simpler right-moving factor \eqref{eq.RmovingCorr}) in terms of a correlation function of four contours of $\partial X$ fields and two insertions of $G$ fields
\begin{align} \label{eq.Aaa11def2}
    A_{m,n}^{(1)(1)} \sim \oint_{\infty}\frac{dt_5}{2\pi i} \oint_{\infty}\frac{dt_6}{2\pi i}&\oint_{-a}\frac{dt_4}{2\pi i}\oint_{-a} \frac{dt_3}{2\pi i} \frac{z_5(t_5)^{m} z_6(t_6)^{n}}{z_4(t_4)^{m} z_3(t_3)^{n}} \nonumber\\
    &\quad\times \langle \partial X_{D\Dd}(t_6) \partial X_{C\Cd}(t_5) G^+_-(t_2) G^-_+(t_1) \partial X_{B\dot{B}}(t_4) \partial X_{A\dot{A}}(t_3) \rangle \ .
\end{align}
This covering space correlator can be computed straightforwardly by breaking the supercurrents into their free-field boson and fermion representation given in \eqref{eq.Gfree} which for the cases we need are
\begin{equation} \label{eq.GpsiX}
    G^+_-(t_2) = \psi^{+E}(t_2)\partial X_{E-}(t_2) \quad,\quad G^-_+(t_1) = \psi^{-F}(t_1)\partial X_{F+}(t_1) \ .
\end{equation}
The correlator in \eqref{eq.Aaa11def2} is then given by
\begin{align} \label{XXGGXXcorrelator}
    \langle& \partial X_{D\Dd}(t_6) \partial X_{C\Cd}(t_5) G^+_-(t_2) G^-_+(t_1) \partial X_{B\dot{B}}(t_4) \partial X_{A\dot{A}}(t_3) \rangle = \frac{1}{t_2-t_1}\bigg[ \nonumber\\
    &\qquad\ \ \: \frac{\epsilon_{CD}\epsilon_{\Cd\Dd}}{(t_6-t_5)^2}\bigg( \frac{2\epsilon_{AB}\epsilon_{\Ad\Bd}}{(t_2-t_1)^2(t_4-t_3)^2} + \frac{\epsilon_{-\Bd}\epsilon_{+\Ad} \epsilon_{AB}}{(t_2-t_4)^2(t_1-t_3)^2} - \frac{\epsilon_{+\Bd}\epsilon_{-\Ad} \epsilon_{AB}}{(t_2-t_3)^2(t_1-t_4)^2} \bigg) \nonumber\\
    &\quad\ \ +\frac{\epsilon_{\Dd-}}{(t_6-t_2)^2} \bigg( \frac{\epsilon_{AB}\epsilon_{\Ad\Bd} \epsilon_{\Cd+}\epsilon_{CD}}{(t_5-t_1)^2(t_4-t_3)^2} + \frac{\epsilon_{BC}\epsilon_{\Bd\Cd} \epsilon_{\Ad+}\epsilon_{A,D}}{(t_5-t_4)^2(t_1-t_3)^2} + \frac{\epsilon_{AC}\epsilon_{\Ad\Cd}\epsilon_{\Bd+}\epsilon_{BD}}{(t_5-t_3)^2(t_1-t_4)^2} \bigg) \nonumber\\
    &\quad\ \ +\frac{-\epsilon_{\Dd+}}{(t_6-t_1)^2} \bigg( \frac{\epsilon_{AB}\epsilon_{\Ad\Bd} \epsilon_{\Cd-}\epsilon_{CD}}{(t_5-t_2)^2(t_4-t_3)^2} + \frac{\epsilon_{BC}\epsilon_{\Bd\Cd} \epsilon_{\Ad-}\epsilon_{AD}}{(t_5-t_4)^2(t_2-t_3)^2} + \frac{\epsilon_{AC}\epsilon_{\Ad\Cd}\epsilon_{\Bd-}\epsilon_{BD}}{(t_5-t_3)^2(t_2-t_4)^2} \bigg) \nonumber\\
    &\quad\ \ +\frac{\epsilon_{BD}\epsilon_{\Bd\Dd}}{(t_6-t_4)^2}\bigg( \frac{2\epsilon_{AC}\epsilon_{\Ad\Cd}}{(t_2-t_1)^2(t_5-t_3)^2} + \frac{\epsilon_{\Ad+}\epsilon_{\Cd-}\epsilon_{AC}}{(t_5-t_2)^2(t_1-t_3)^2} - \frac{\epsilon_{\Ad-}\epsilon_{\Cd+}\epsilon_{AC}}{(t_5-t_1)^2(t_2-t_3)^2} \bigg) \nonumber\\
    &\quad\ \ +\frac{\epsilon_{AD}\epsilon_{\Ad\Dd}}{(t_6-t_3)^2}\bigg( \frac{2\epsilon_{BC}\epsilon_{\Bd\Cd}}{(t_2-t_1)^2(t_5-t_4)^2} + \frac{\epsilon_{\Bd+}\epsilon_{\Cd-}\epsilon_{BC}}{(t_5-t_2)^2(t_1-t_4)^2} - \frac{\epsilon_{\Bd-}\epsilon_{\Cd+}\epsilon_{BC}}{(t_5-t_1)^2(t_2-t_4)^2} \bigg)  \bigg] \ .
\end{align}
The four contour integrals in \eqref{eq.Aaa11def2} can then be straightforwardly evaluated, yielding the amplitude $A^{(1)(1)}_{m,n}$. The equivalent amplitude with both the initial and final states being excited on copy 2 rather than copy 1 is then trivially obtained since by symmetry $A^{(2)(2)}_{m,n}=A^{(1)(1)}_{m,n}$. This then leaves the computation of the amplitudes with initial and final state excitations on different copies, \textit{i.e.} $A^{(2)(1)}_{m,n}=A^{(1)(2)}_{m,n}$ (again by symmetry). As explained in \cite{Hampton:2018ygz} these correlators can be obtained by moving one of the deformation operators once around the doubly-covered cylinder on which the amplitude is defined. This has the effect of interchanging the two copies that are being twisted together. By writing the computed amplitude $A^{(1)(1)}_{m,n}$ back in terms of the coordinates on the cylinder $(w_2=w,w_1=0)$ and mapping the insertion of the second deformation operator using $w\to w+2\pi i$ we obtain $A^{(2)(1)}_{m,n}$. The sum of amplitudes
\begin{equation} \label{eq.Asymm}
    A_{m,n}(w_2,w_1) = \frac12 \sum_{i,j=1}^{2} A^{(i)(j)}_{m,n}(w_2,w_1) \ ,
\end{equation}
is then the copy-symmetric amplitude required for the lift, where the factor of $\frac12$ comes from the normalisation over copy indices in the state \eqref{eq.aaState}. We then obtain the integrated amplitude $X(T)$ defined in \eqref{eq.Xdef} by integrating over the insertion points $w_1,w_2$. As argued in Section~\ref{sec:LiftForm}, the integrated amplitude can be written as a sum of three contour integrals $I_{C_1},I_{C_2},I_{C_3}$ defined in \eqref{eq.IC123defs} with $I_{C_3}$ having a vanishing contribution. It can also be shown that for the states we consider $\lim_{T\to\infty}I_{C_1}(T)=\lim_{T\to\infty}I_{C_2}(T)$ and so this step can be reduced to essentially one independent contour integral leaving
\begin{equation} \label{eq.aaliftForm}
   E^{(2)}\big(|\alpha\alpha\rangle_{B\Bd A\Ad(m,n)}\big) = \frac{\pi\lambda^2}{\big||\alpha\alpha\rangle_{\!B\Bd A\Ad(m,n)}\big|^2} \lim_{T\to\infty} \int^{2\pi}_0 \!d\sigma\, A_{m,n}(w,0) \coth(\tfrac{\bar w}{2}) \ ,
\end{equation}
where $w= \frac{T}{2} + i \sigma$ and $\wb = \frac{T}{2} - i\sigma$ with $T\to \infty$ on this contour. For generic $SU(2)$ doublet indices on the $\alpha$ modes in the amplitude \eqref{eq.Aaa11def} it is not directly related to the second order lift of a particular state of the form \eqref{eq.aaState}. Only when the final state is the Hermitian conjugate of the initial state (as defined in Appendix~\ref{app.Herm}) can we use the relation \eqref{eq.aaliftForm} to obtain a lift.\\
\\
As an example we give below a matrix of lifts for the particular states $|\alpha\alpha\rangle_{\!++++(m,n)}$:
\\
\begin{align}
\label{eq.aamat}
\tikz[remember picture, baseline=(mat.center)]{\node[inner sep=0](mat){$
\begin{pmatrix}
\frac{\pi^2}{2} & \frac{\pi^2}{2} & \frac{9 \pi^2}{16} & \frac{19 \pi^2}{32} & \frac{5045 \pi^2}{8192} & \frac{10377 \pi^2}{16384} & \frac{42469 \pi^2}{65536} \\\\
\frac{\pi^2}{2} & \frac{19 \pi^2}{32} & \frac{9 \pi^2}{16} & \frac{2569 \pi^2}{4096} & \frac{10775 \pi^2}{16384} & \frac{44469 \pi^2}{65536} & \frac{22771 \pi^2}{32768} \\\\
\frac{9 \pi^2}{16} & \frac{9 \pi^2}{16} & \frac{2697 \pi^2}{4096} & \frac{5001 \pi^2}{8192} & \frac{44205 \pi^2}{65536} & \frac{23061 \pi^2}{32768} & \frac{24275979 \pi^2}{33554432} \\\\
\frac{19 \pi^2}{32} & \frac{2569 \pi^2}{4096} & \frac{5001 \pi^2}{8192} & \frac{92751 \pi^2}{131072} & \frac{42555 \pi^2}{65536} & \frac{47798349 \pi^2}{67108864} & \frac{99368101 \pi^2}{134217728} \\\\
\frac{5045 \pi^2}{8192} & \frac{10775 \pi^2}{16384} & \frac{44205 \pi^2}{65536} & \frac{42555 \pi^2}{65536} & \frac{200592615 \pi^2}{268435456} & \frac{183067725 \pi^2}{268435456} & \frac{1597261925 \pi^2}{2147483648} \\\\
\frac{10377 \pi^2}{16384} & \frac{44469 \pi^2}{65536} & \frac{23061 \pi^2}{32768} & \frac{47798349 \pi^2}{67108864} & \frac{183067725 \pi^2}{268435456} & \frac{3351953103 \pi^2}{4294967296} & \frac{1524951729 \pi^2}{2147483648} \\\\
\frac{42469 \pi^2}{65536} & \frac{22771 \pi^2}{32768} & \frac{24275979 \pi^2}{33554432} & \frac{99368101 \pi^2}{134217728} & \frac{1597261925 \pi^2}{2147483648} & \frac{1524951729 \pi^2}{2147483648} & \frac{444727449481 \pi^2}{549755813888} \\
\end{pmatrix}
$};}
\begin{tikzpicture}[overlay,remember picture]
\draw[blue,thick,->] node[anchor=south west] (nn1) at (mat.north west)
{increasing $n$} (nn1.east) -- (nn1-|mat.north east) ;
\draw[red,thick,->] node[anchor=north west,align=center, inner xsep=0pt] (nn2) at 
(mat.north east)
{\rotatebox{-90}{increasing $m$}} (nn2.south) -- (nn2.south|-mat.south) 
;
\end{tikzpicture}
\end{align}
\\
This lift matrix is symmetric in the mode numbers $m,n$ since the two alpha modes commute. The method for computing lifts used in this paper has the advantage of being able to go to high levels which we demonstrate for the following families.
\begin{figure}[H]
    \centering
    \vspace{-5pt}\fbox{\includegraphics[width=0.45\textwidth
    ]{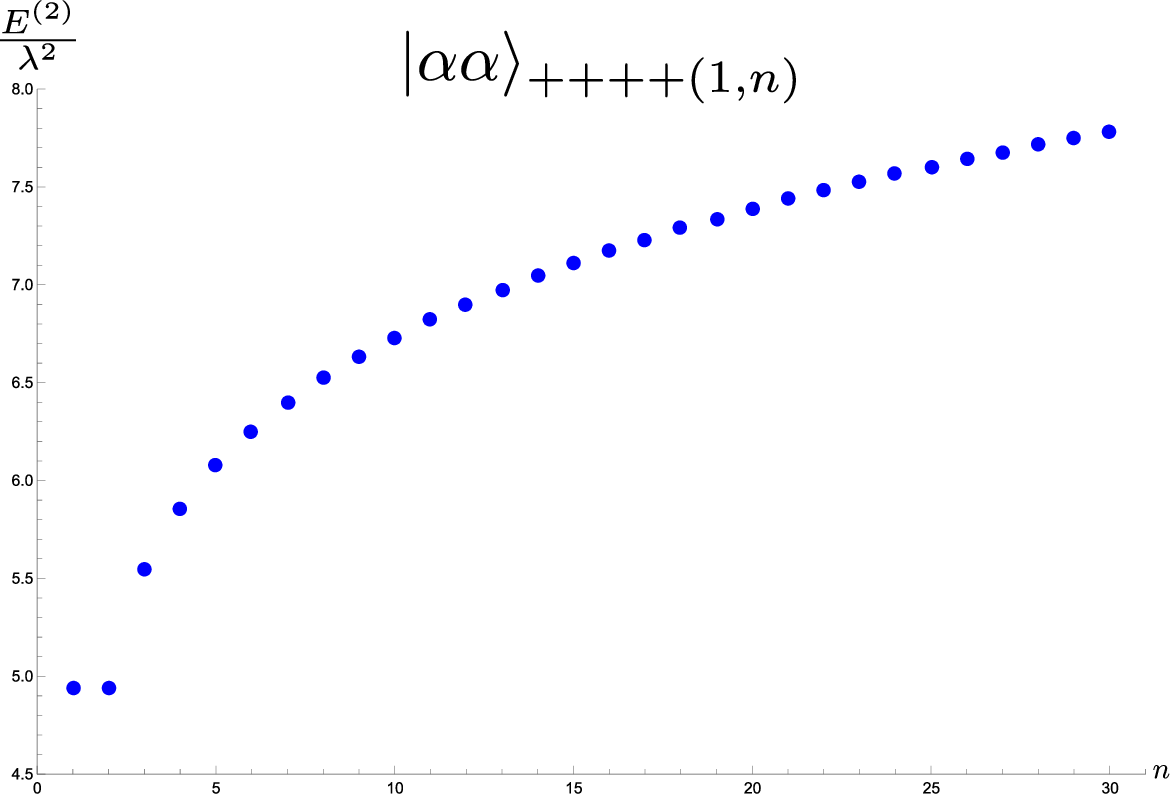}}
    \caption{Plot of the lifts $E^{(2)}(|\alpha\alpha\rangle_{\!++++(1,n)})/\lambda^2$ for varying $n$. The plot fits to the curve: $-0.0974563\; n+1.28434 \sqrt{n}+3.62389$.\label{fig:aa1}}
\end{figure}
\begin{figure}[H]
    \centering
    \vspace{-15pt}\fbox{\includegraphics[width=0.45\textwidth
    ]{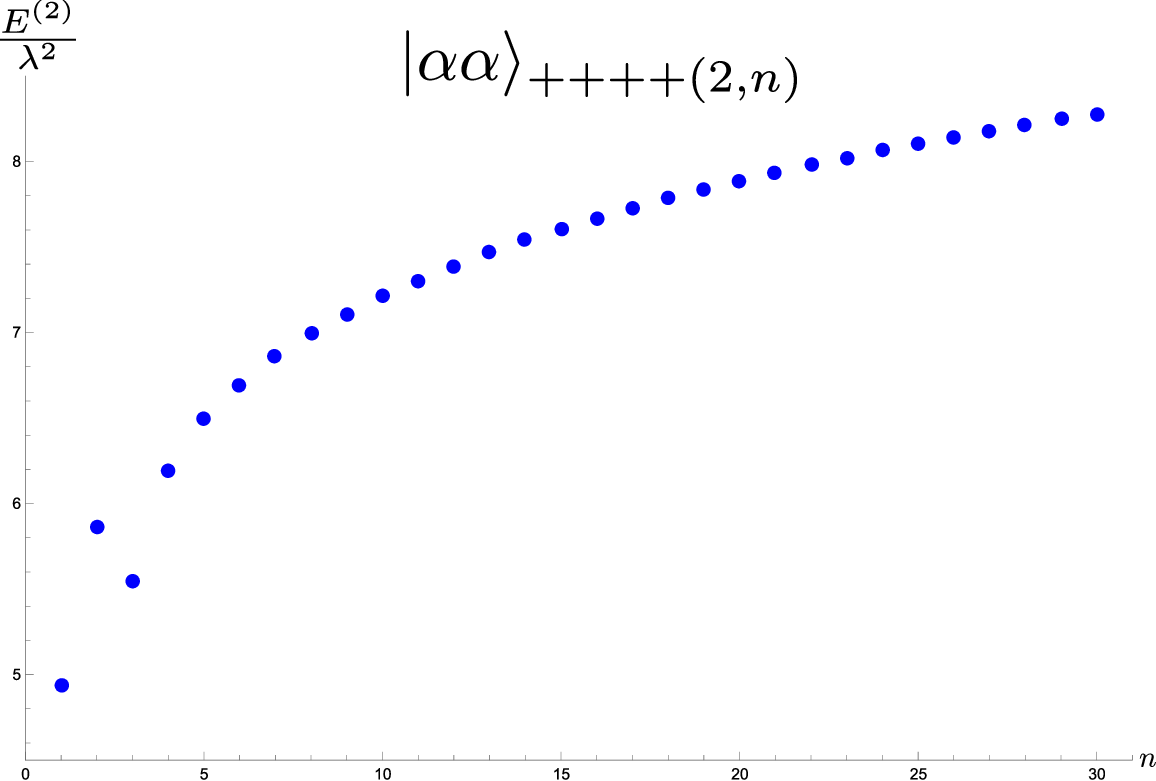}}
    \caption{Plot of the lifts $E^{(2)}(|\alpha\alpha\rangle_{\!++++(2,n)})/\lambda^2$ for varying $n$. The plot fits to the curve:  $-0.119293\; n+1.47589 \sqrt{n}+3.71017$.}
\end{figure}
\begin{figure}[H]
    \centering
    \vspace{-15pt}\fbox{\includegraphics[width=0.45\textwidth
    ]{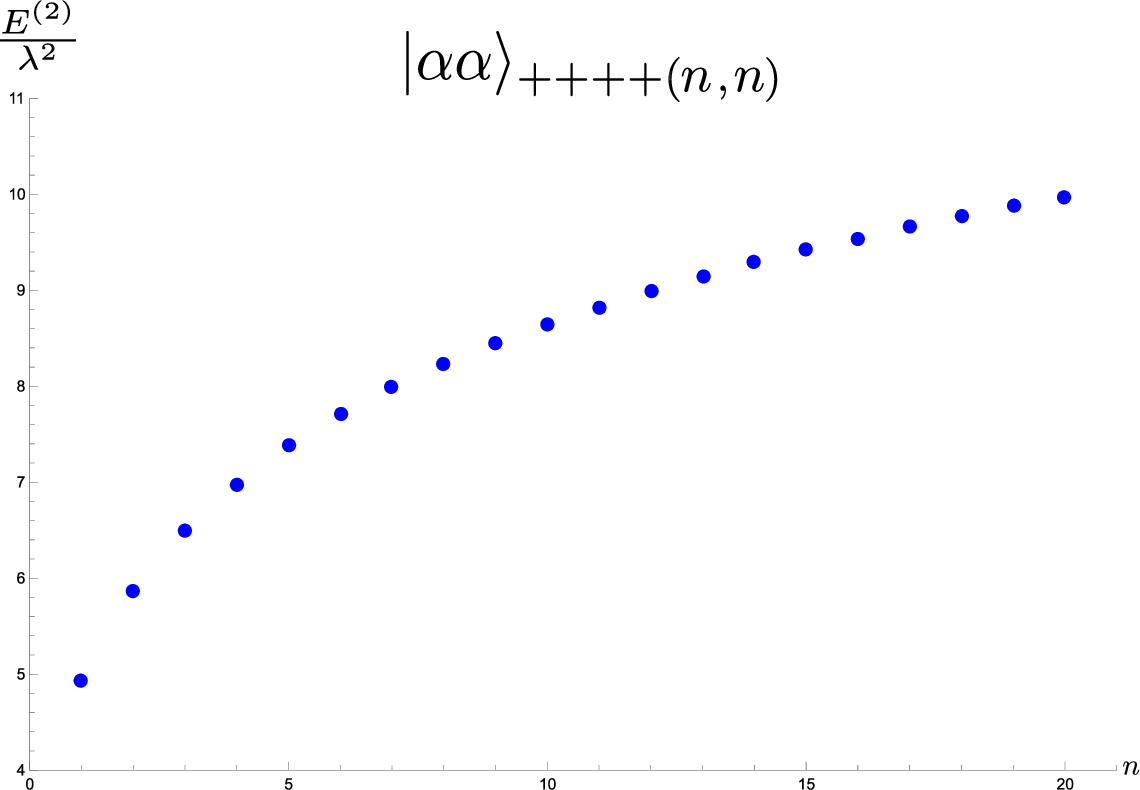}}
    \caption{Plot of the lifts $E^{(2)}(|\alpha\alpha\rangle_{\!++++(n,n)})/\lambda^2$ for varying $n$. The plot fits to the curve:  $-0.211721\; n+2.57926 \sqrt{n}+2.62963$.}
\end{figure}

\begin{table}[H]
\resizebox{\columnwidth}{!}{
\begin{tabular}{|c|c|c|c|}
\hline
&&&\\[-1em]
$n$ & $E^{(2)}\big(|\alpha\alpha\rangle_{\!++++(1,n)}\big)/\lambda^2$ & $E^{(2)}(|\alpha\alpha\rangle_{\!++++(2,n)})/\lambda^2$ & $E^{(2)}(|\alpha\alpha\rangle_{\!++++(n,n)})/\lambda^2$ \\[3pt]
\hline
&&&\\[-1em]
1 & $\frac{\pi^2}{2}$ & $\frac{\pi^2}{2}$ & $\frac{\pi^2}{2}$ \\[3pt]
\hline
&&&\\[-1em]
2 & $\frac{\pi^2}{2}$ & $\frac{19 \pi^2}{32}$ & $\frac{19 \pi^2}{32}$ \\[3pt]
\hline
&&&\\[-1em]
3 & $\frac{9 \pi^2}{16}$ & $\frac{9 \pi^2}{16}$ & $\frac{2697 \pi^2}{4096}$ \\[3pt]
\hline
&&&\\[-1em]
4 & $\frac{19 \pi^2}{32}$ & $\frac{2569 \pi^2}{4096}$ & $\frac{92751 \pi^2}{131072}$ \\[3pt]
\hline
&&&\\[-1em]
5 & $\frac{5045 \pi^2}{8192}$ & $\frac{10775 \pi^2}{16384}$ & $\frac{200592615 \pi^2}{268435456}$ \\[3pt]
\hline
&&&\\[-1em]
6 & $\frac{10377 \pi^2}{16384}$ & $\frac{44469 \pi^2}{65536}$ & $\frac{3351953103 \pi^2}{4294967296}$ \\[3pt]
\hline
&&&\\[-1em]
7 & $\frac{42469 \pi^2}{65536}$ & $\frac{22771 \pi^2}{32768}$ & $\frac{444727449481 \pi^2}{549755813888}$ \\[3pt]
\hline
&&&\\[-1em]
8 & $\frac{86599 \pi^2}{131072}$ & $\frac{95104151 \pi^2}{134217728}$ & $\frac{29342363983399 \pi^2}{35184372088832}$ \\[3pt]
\hline
&&&\\[-1em]
9 & $\frac{360715167 \pi^2}{536870912}$ & $\frac{773523243 \pi^2}{1073741824}$ & $\frac{987157918885637403 \pi^2}{1152921504606846976}$ \\[3pt]
\hline
&&&\\[-1em]
10 & $\frac{732197015 \pi^2}{1073741824}$ & $\frac{6278009315 \pi^2}{8589934592}$ & $\frac{16164598479554313885 \pi^2}{18446744073709551616}$ \\[3pt]
\hline
&&&\\[-1em]
11 & $\frac{5935693181 \pi^2}{8589934592}$ & $\frac{6358760023 \pi^2}{8589934592}$ & $\frac{2112174651194076415551 \pi^2}{2361183241434822606848}$ \\[3pt]
\hline
&&&\\[-1em]
12 & $\frac{12014401671 \pi^2}{17179869184}$ & $\frac{1646648430939 \pi^2}{2199023255552}$ & $\frac{68855070668333647215123 \pi^2}{75557863725914323419136}$ \\[3pt]
\hline
&&&\\[-1em]
13 & $\frac{3109456016173 \pi^2}{4398046511104}$ & $\frac{6655660569571 \pi^2}{8796093022208}$ & $\frac{143409167147170367122924179 \pi^2}{154742504910672534362390528}$ \\[3pt]
\hline
&&&\\[-1em]
14 & $\frac{6281601504073 \pi^2}{8796093022208}$ & $\frac{13439095533577 \pi^2}{17592186044416}$ & $\frac{2330135028016340441236500951 \pi^2}{2475880078570760549798248448}$ \\[3pt]
\hline
&&&\\[-1em]
15 & $\frac{12680206592715 \pi^2}{17592186044416}$ & $\frac{6779007700515 \pi^2}{8796093022208}$ & $\frac{302511018713745232822331768265 \pi^2}{316912650057057350374175801344}$ \\[3pt]
\hline
&&&\\[-1em]
16 & $\frac{25579799605231 \pi^2}{35184372088832}$ & $\frac{223957295729906611 \pi^2}{288230376151711744}$ & $\frac{39232078561011585629574541281111 \pi^2}{40564819207303340847894502572032}$ \\[3pt]
\hline
&&&\\[-1em]
17 & $\frac{1689931464780479363 \pi^2}{2305843009213693952}$ & $\frac{3610722048060129083 \pi^2}{4611686018427387904}$ & $\frac{20820990981442283120128351763273228067 \pi^2}{21267647932558653966460912964485513216}$ \\[3pt]
\hline
&&&\\[-1em]
18 & $\frac{3405431755536511923 \pi^2}{4611686018427387904}$ & $\frac{29092589394600843909 \pi^2}{36893488147419103232}$ & $\frac{336946377389953804927663882067384419377 \pi^2}{340282366920938463463374607431768211456}$ \\[3pt]
\hline
&&&\\[-1em]
19 & $\frac{27437286666189088727 \pi^2}{36893488147419103232}$ & $\frac{29288337033505778419 \pi^2}{36893488147419103232}$ & $\frac{43591348921018738029213125002844429996659 \pi^2}{43556142965880123323311949751266331066368}$ \\[3pt]
\hline
&&&\\[-1em]
20 & $\frac{55242956792501362205 \pi^2}{73786976294838206464}$ & $\frac{7545386463175746715115 \pi^2}{9444732965739290427392}$ & $\frac{1408978149313722626429865625800090243241265 \pi^2}{1393796574908163946345982392040522594123776}$ \\[3pt]
\hline
\end{tabular}
}\caption{We give the values of lifts of the state $|\alpha\alpha\rangle_{++++(m,n)}$ for two families where one mode has a small fixed mode number and the other mode is varied, and one family where the two modes are taken to be equal. We choose to present the first twenty lifts in each of these families.\label{aatab}}
\end{table}

\subsection{Lifting of $|\alpha d\rangle_{\!B\Bd(m,s)}^{\alpha A}$} \label{ssec.Liftad}

Next we consider the lift of states of the form
\begin{equation} \label{eq.adState}
    |\alpha d\rangle_{\!B\Bd(m,s)}^{\alpha A} \equiv \frac{1}{\sqrt{2}} \Big(\alpha^{(1)}_{B\Bd,-m} d^{(1)\alpha A}_{-s} + \alpha^{(2)}_{B\Bd,-m} d^{(2)\alpha A}_{-s}\Big) \NSok{1}\NSok{2} \ ,
\end{equation}
where the norm is given by
\begin{align} \label{eq.adNorm}
    {}_{(m,s)C\Cd}^{\qquad\delta D}\langle\alpha d|\alpha d\rangle_{\!B\Bd(m,s)}^{\alpha A} = m\, \epsilon_{BC} \epsilon_{\Bd\Cd} \epsilon^{\alpha\delta} \epsilon^{AD} H[m-1] H\big[s-\tfrac12\big] \ ,
\end{align}
with appropriate choices of $SU(2)$ indices as per Appendix~\ref{app.Herm}. The computation of the lift requires the following left-moving amplitude
\begin{equation} \label{eq.Aad11def}
    A_{m,s}^{(1)(1)}(w_2,w_1) \equiv \langle\NSo|\big(d^{(1)\delta D}_{s}\alpha^{(1)}_{C\Cd,m}\big)\, \Big(G^+_{-,-\frac12}\sigma^-\Big)(w_2) \Big(G^-_{+,-\frac12}\sigma^+\Big)(w_1)\, \big(\alpha^{(1)}_{B\Bd,-m}d^{(1)\alpha A}_{-s}\big) |\NSo\rangle \ ,
\end{equation}
where we have chosen all of the modes to act on copy 1. In order to have a non-vanishing state we clearly require
\begin{equation}
    n>0 \quad\text{and}\quad s\geq\frac12\ ,
\end{equation}
since the $d$ and $\alpha$ modes commute and so neither should annihilate the NS vacuum. In mapping to the covering space and removing the spin fields that appear there via spectral flow transformations, the initial and final states transform similarly to the $|\alpha\alpha\rangle$ states in \eqref{eq.aatransi} and \eqref{eq.aatransf}, however, the $d$ modes gain extra factors due to their non-zero charge under $J^3_0$. The initial state therefore transforms as
\begin{align} \label{eq.adtransi}
    \big(\alpha^{(1)}_{B\dot B,-m} d^{(1)\alpha A}_{-s}\big)_{-\infty} &= i \oint_{-\infty}\frac{dw_4}{2\pi i}\oint_{-\infty} \frac{dw_3}{2\pi i}\, e^{-m w_4}e^{-s w_3} \, \partial X_{B\dot{B}}(w_4) \psi^{\alpha A}(w_3) \nonumber\\
    &= i \oint_{0}\frac{dz_4}{2\pi i}\oint_{0} \frac{dz_3}{2\pi i} \,z_4^{-m}z_3^{-s-\frac12} \partial X_{B\dot{B}}(z_4) \psi^{\alpha A}(z_3) \nonumber\\
    &= i \oint_{-a}\frac{dt_4}{2\pi i}\oint_{-a} \frac{dt_3}{2\pi i} \bigg(\frac{dz_3}{dt_3}\bigg)^{\!\frac12} \, \frac{\partial X_{B\dot{B}}(t_4) \psi^{\alpha A}(t_3)}{z_4(t_4)^{m} z_3(t_3)^{s+\frac12}} \nonumber\\
    &\longrightarrow i \oint_{-a}\frac{dt_4}{2\pi i}\oint_{-a} \frac{dt_3}{2\pi i} \bigg(\frac{dz_3}{dt_3}\bigg)^{\!\frac12}\bigg(\frac{t_3-t_1}{t_3-t_2}\bigg)^{\!q_{\alpha}} \, \frac{\partial X_{B\dot{B}}(t_4) \psi^{\alpha A}(t_3)}{z_4(t_4)^{m} z_3(t_3)^{s+\frac12}} \ ,
\end{align}
where the spectral flow transformations were made in the final line with $q_{\alpha}$ being the eigenvalue of $J^3_0$ for the state created by the fermion mode $d^{\alpha A}_{-s}$. Likewise the final state transforms as
\begin{align} \label{eq.adtransf}
    \big(d^{(1)\delta D}_{s}\alpha^{(1)}_{C\dot C,m}\big)_{\infty} &= i \oint_{\infty}\frac{dw_5}{2\pi i}\oint_{\infty} \frac{dw_6}{2\pi i}\, e^{m w_5} e^{s w_6} \, \psi^{\delta D}(w_6) \partial X_{C\Cd}(w_5) \nonumber\\
    &= i \oint_{\infty}\frac{dz_5}{2\pi i}\oint_{\infty} \frac{dz_6}{2\pi i}\, z_5^{m} z_6^{s-\frac12} \, \psi^{\delta D}(z_6) \partial X_{C\Cd}(z_5) \nonumber\\
    &= i \oint_{\infty}\frac{dt_5}{2\pi i}\oint_{\infty} \frac{dt_6}{2\pi i}\, z_5(t_5)^{m} z_6(t_6)^{s-\frac12} \bigg(\frac{dz_6}{dt_6}\bigg)^{\!\frac12}\, \psi^{\delta D}(t_6) \partial X_{C\Cd}(t_5) \nonumber\\
    &\longrightarrow i \oint_{\infty}\frac{dt_5}{2\pi i}\oint_{\infty} \frac{dt_6}{2\pi i} \,z_5(t_5)^{m} z_6(t_6)^{s-\frac12} \bigg(\frac{dz_6}{dt_6}\bigg)^{\!\frac12}\bigg(\frac{t_6-t_1}{t_6-t_2}\bigg)^{\!q_{\delta}} \psi^{\delta D}(t_6) \partial X_{C\Cd}(t_5) \ .
\end{align}
This leaves the amplitude \eqref{eq.Aad11def} (up to overall factors) as a correlation function of two contours of $\partial X$ fields, two contours of $\psi$ fields and two insertions of $G$ fields
\begin{align} \label{eq.Aad11def2}
    A_{m,s}^{(1)(1)} \sim &\oint_{\infty}\frac{dt_5}{2\pi i} \oint_{\infty}\frac{dt_6}{2\pi i}\oint_{-a}\frac{dt_4}{2\pi i}\oint_{-a} \frac{dt_3}{2\pi i} \frac{z_5(t_5)^{m} z_6(t_6)^{s-\frac12}}{z_4(t_4)^{m} z_3(t_3)^{s+\frac12}} \bigg(\frac{dz_6}{dt_6}\bigg)^{\!\frac12}\bigg(\frac{dz_3}{dt_3}\bigg)^{\!\frac12}\bigg(\frac{t_6-t_1}{t_6-t_2}\bigg)^{\!q_{\delta}}\nonumber\\
    &\quad\times \bigg(\frac{t_3-t_1}{t_3-t_2}\bigg)^{\!q_{\alpha}} \langle \psi^{\delta D}(t_6) \partial X_{C\Cd}(t_5) G^+_-(t_2) G^-_+(t_1) \partial X_{B\dot{B}}(t_4) \psi^{\alpha A}(t_3) \rangle \ .
\end{align}
This covering space correlator can be computed straightforwardly by breaking the supercurrents into the free-field bosons and fermions via \eqref{eq.GpsiX}, yielding
\begin{align}
    \langle& \psi^{\delta D}(t_6) \partial X_{C\Cd}(t_5) G^+_-(t_2) G^-_+(t_1) \partial X_{B\dot{B}}(t_4) \psi^{\alpha A}(t_3) \rangle = \frac{\epsilon_{\Cd-}\epsilon_{+\Bd} }{(t_5-t_2)^2(t_1-t_4)^2}\Bigg[ \frac{\epsilon^{\delta+}\epsilon^{-\alpha}\delta^{D}_{C}\delta^{A}_{B}}{(t_6-t_2)(t_1-t_3)} \nonumber\\
    &\quad- \frac{\epsilon^{\delta-}\epsilon^{+\alpha}\delta^{D}_{B}\delta^{A}_{C}}{(t_6-t_1)(t_2-t_3)} + \frac{\epsilon^{\delta\alpha}\epsilon^{AD}\epsilon_{CB}}{(t_6-t_3)(t_2-t_1)}\Bigg]
    + \frac{\epsilon_{\Cd+}\epsilon_{2\Bd}}{(t_5-t_1)^2(t_2-t_4)^2}\Bigg[ \frac{\epsilon^{\delta+}\epsilon^{-\alpha}\delta^{A}_{C}\delta^{D}_{B}}{(t_6-t_2)(t_1-t_3)} \nonumber\\
    &\quad- \frac{\epsilon^{\delta-}\epsilon^{+\alpha}\delta^{D}_{C}\delta^{A}_{B}}{(t_6-t_1)(t_2-t_3)} - \frac{\epsilon^{\delta\alpha}\epsilon^{AD}\epsilon_{CB}}{(t_6-t_3)(t_2-t_1)}\Bigg]
    + \frac{\epsilon_{CB}\epsilon_{\Cd\Bd}}{(t_5-t_4)^2(t_2-t_1)^2}\Bigg[ \frac{\epsilon^{\delta+}\epsilon^{-\alpha}\epsilon^{AD}}{(t_6-t_2)(t_1-t_3)} \nonumber\\
    &\quad+ \frac{\epsilon^{\delta-}\epsilon^{+\alpha}\epsilon^{AD}}{(t_6-t_1)(t_2-t_3)} + \frac{2\epsilon^{\delta\alpha}\epsilon^{AD}}{(t_6-t_3)(t_2-t_1)}\Bigg] \ .
\end{align}
The remainder of the lifting computation is exactly as described in Section~\ref{ssec.Liftaa} with the final lift given by
\begin{equation} \label{eq.adliftForm}
   E^{(2)}\big(|\alpha d\rangle^{\alpha A}_{B\Bd(m,s)}\big) = \frac{\pi\lambda^2}{\big||\alpha d\rangle^{\alpha A}_{\!B\Bd(m,s)}\big|^2} \lim_{T\to\infty} \int^{2\pi}_0 \!d\sigma\, A_{m,s}(w,0) \coth(\tfrac{\bar w}{2}) \ ,
\end{equation}
where $w= \frac{T}{2} + i \sigma$ and $\wb = \frac{T}{2} - i\sigma$ with $T\to \infty$ on this contour and $A_{m,s}(w_2,w_1)$ is the copy-symmetrised amplitude defined analogously to \eqref{eq.Asymm}. For generic $SU(2)$ doublet indices on the $\alpha$ and $d$ modes in the amplitude \eqref{eq.Aad11def} it is not directly related to the second order lift of a particular state of the form \eqref{eq.adState}. Only when the final state is the Hermitian conjugate of the initial state (as defined in Appendix~\ref{app.Herm}) can we use the relation \eqref{eq.adliftForm} to obtain a lift.\\
\\
As an example, the matrix in mode numbers $m,s$ of the lifts of the states $|\alpha d\rangle_{\!++(m,s)}^{--}$ are given below:
\\
\begin{align}
\label{eq.admat}
\tikz[remember picture, baseline=(mat.center)]{\node[inner sep=0](mat){$
\begin{pmatrix}
\frac{\pi^2}{2} & \frac{\pi^2}{2} & \frac{35 \pi^2}{64} & \frac{37 \pi^2}{64} & \frac{9859 \pi^2}{16384} & \frac{10171 \pi^2}{16384} & \frac{333847 \pi^2}{524288} \\\\
\frac{\pi^2}{2} & \frac{19 \pi^2}{32} & \frac{19 \pi^2}{32} & \frac{1283 \pi^2}{2048} & \frac{10663 \pi^2}{16384} & \frac{175711 \pi^2}{262144} & \frac{179969 \pi^2}{262144} \\\\
\frac{33 \pi^2}{64} & \frac{39 \pi^2}{64} & \frac{2697 \pi^2}{4096} & \frac{2697 \pi^2}{4096} & \frac{358269 \pi^2}{524288} & \frac{368643 \pi^2}{524288} & \frac{48290217 \pi^2}{67108864} \\\\
\frac{17 \pi^2}{32} & \frac{2563 \pi^2}{4096} & \frac{5525 \pi^2}{8192} & \frac{92751 \pi^2}{131072} & \frac{92751 \pi^2}{131072} & \frac{48833787 \pi^2}{67108864} & \frac{99898097 \pi^2}{134217728} \\\\
\frac{8935 \pi^2}{16384} & \frac{10495 \pi^2}{16384} & \frac{361355 \pi^2}{524288} & \frac{378645 \pi^2}{524288} & \frac{200592615 \pi^2}{268435456} & \frac{200592615 \pi^2}{268435456} & \frac{6562949705 \pi^2}{8589934592} \\\\
\frac{9141 \pi^2}{16384} & \frac{171381 \pi^2}{262144} & \frac{184155 \pi^2}{262144} & \frac{24675753 \pi^2}{33554432} & \frac{204097593 \pi^2}{268435456} & \frac{3351953103 \pi^2}{4294967296} & \frac{3351953103 \pi^2}{4294967296} \\\\
\frac{298417 \pi^2}{524288} & \frac{348943 \pi^2}{524288} & \frac{47941229 \pi^2}{67108864} & \frac{50147797 \pi^2}{67108864} & \frac{6632510507 \pi^2}{8589934592} & \frac{6804589421 \pi^2}{8589934592} & \frac{444727449481 \pi^2}{549755813888} \\
\end{pmatrix}
$};}
\begin{tikzpicture}[overlay,remember picture]
\draw[blue,thick,->] node[anchor=south west] (nn1) at (mat.north west)
{increasing $s$} (nn1.east) -- (nn1-|mat.north east) ;
\draw[red,thick,->] node[anchor=north west,align=center, inner xsep=0pt] (nn2) at 
(mat.north east)
{\rotatebox{-90}{increasing $m$}} (nn2.south) -- (nn2.south|-mat.south) 
;
\end{tikzpicture}
\end{align}%
\begin{figure}[H]
    \centering
    \fbox{\includegraphics[scale=0.51]{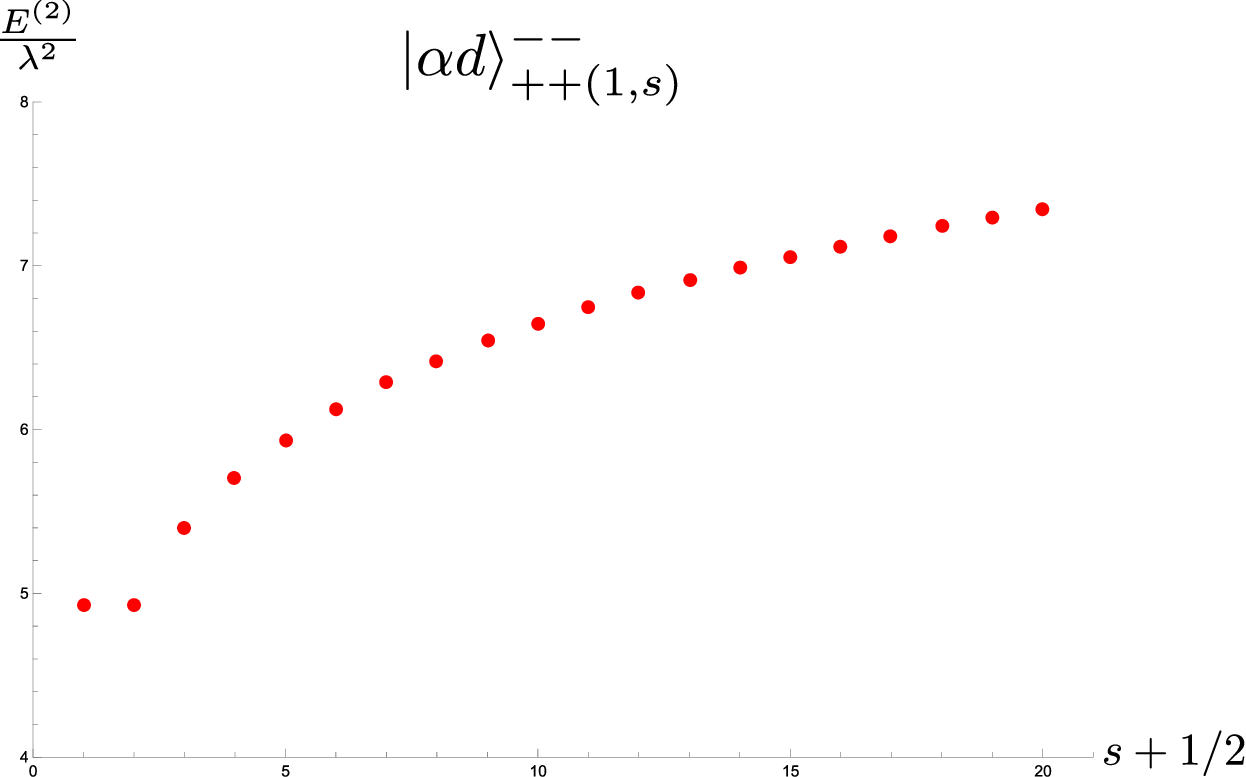}}
    \caption{Plot of the lifts $E^{(2)}\big(|\alpha d\rangle_{\!++(1,s)}^{--}\big)/\lambda^2$ for varying $s$. The plot fits to the curve: $-0.0187122(s+1/2)+0.590457 \sqrt{s+1/2}+4.20151$.}
    \label{fig:6}
\end{figure}
\begin{figure}[H]
    \centering
    \fbox{\includegraphics[scale=0.55]{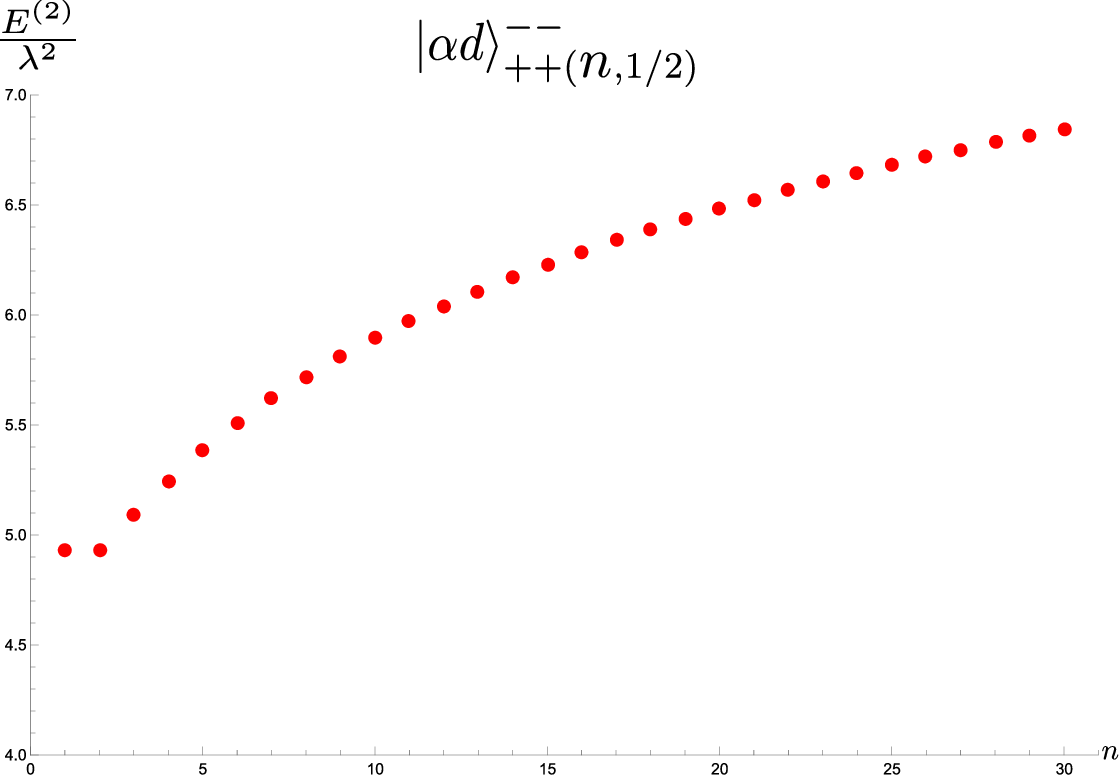}}
    \caption{Plot of the lifts $E^{(2)}\big(|\alpha d\rangle_{\!++(n,1/2)}^{--}\big)/\lambda^2$ for varying $n$. The plot fits to the curve: $-0.0909845\; n+1.25023 \sqrt{n}+3.58201$.}
    \label{fig:7}
\end{figure}
\noindent Below we do not list the diagonal elements of the lift matrix for large mode numbers of $|\alpha d\rangle_{\!++(m,s)}^{--}$ because there are relations between the diagonal elements of the lift matrix of states $|\alpha\alpha\rangle_{\!++++(m,n)}$ and $|\alpha d\rangle_{\!++(m,s)}^{--}$. These relations force the diagonal elements of the lift matrix of $|\alpha d\rangle_{\!++(m,s)}^{--}$ to be the same as those of $|\alpha\alpha\rangle_{\!++++(m,n)}$. We study these relations in Section~\ref{sec.checks}.
\begin{table}[H]
\centering
\begin{tabular}{|c|c|c|c|c|}
\hline
&&&&\\[-1em]
 $s$& $E^{(2)}(|\alpha d\rangle_{\!++(1,s)}^{--})/\lambda^2$ & & $m$ & $E^{(2)}(|\alpha d\rangle_{\!++(m,1/2)}^{--})/\lambda^2$ \\[3pt]
\hline
&&&&\\[-1em]
1/2 & $\frac{\pi^2}{2}$ && 1 & $\frac{\pi^2}{2}$ \\[3pt]
\hline
&&&&\\[-1em]
3/2 & $\frac{\pi^2}{2}$ && 2 & $\frac{\pi^2}{2}$ \\[3pt]
\hline
&&&&\\[-1em]
5/2 & $\frac{35 \pi^2}{64}$ && 3 & $\frac{33 \pi^2}{64}$ \\[3pt]
\hline
&&&&\\[-1em]
7/2 & $\frac{37 \pi^2}{64}$ && 4 & $\frac{17 \pi^2}{32}$ \\[3pt]
\hline
&&&&\\[-1em]
9/2 & $\frac{9859 \pi^2}{16384}$ && 5 & $\frac{8935 \pi^2}{16384}$ \\[3pt]
\hline
&&&&\\[-1em]
11/2 & $\frac{10171 \pi^2}{16384}$ && 6 & $\frac{9141 \pi^2}{16384}$ \\[3pt]
\hline
&&&&\\[-1em]
13/2 & $\frac{333847 \pi^2}{524288}$ && 7 & $\frac{298417 \pi^2}{524288}$ \\[3pt]
\hline
&&&&\\[-1em]
15/2 & $\frac{341065 \pi^2}{524288}$ && 8 & $\frac{75937 \pi^2}{131072}$ \\[3pt]
\hline
&&&&\\[-1em]
17/2 & $\frac{711494891 \pi^2}{1073741824}$ && 9 & $\frac{632011347 \pi^2}{1073741824}$ \\[3pt]
\hline
&&&&\\[-1em]
19/2 & $\frac{723089227 \pi^2}{1073741824}$ && 10 & $\frac{641119135 \pi^2}{1073741824}$ \\[3pt]
\hline
&&&&\\[-1em]
21/2 & $\frac{23473859357 \pi^2}{34359738368}$ && 11 & $\frac{20784725687 \pi^2}{34359738368}$ \\[3pt]
\hline
&&&&\\[-1em]
23/2 & $\frac{23779258907 \pi^2}{34359738368}$ && 12 & $\frac{10517135061 \pi^2}{17179869184}$ \\[3pt]
\hline
&&&&\\[-1em]
25/2 & $\frac{6159330683695 \pi^2}{8796093022208}$ && 13 & $\frac{5444354499883 \pi^2}{8796093022208}$ \\[3pt]
\hline
&&&&\\[-1em]
27/2 & $\frac{6225785037127 \pi^2}{8796093022208}$ && 14 & $\frac{5500170966829 \pi^2}{8796093022208}$ \\[3pt]
\hline
&&&&\\[-1em]
29/2 & $\frac{201203440824383 \pi^2}{281474976710656}$ && 15 & $\frac{177685335597585 \pi^2}{281474976710656}$ \\[3pt]
\hline
&&&&\\[-1em]
31/2 & $\frac{203052922651009 \pi^2}{281474976710656}$  && 16 & $\frac{22408851223553 \pi^2}{35184372088832}$ \\[3pt]
\hline
&&&&\\[-1em]
33/2 & $\frac{3355269041672768731 \pi^2}{4611686018427387904}$ && 17 & $\frac{2961766835461728811 \pi^2}{4611686018427387904}$ \\[3pt]
\hline
&&&&\\[-1em]
35/2 & $\frac{3382080970269418075 \pi^2}{4611686018427387904}$ && 18 & $\frac{2985117620728822659 \pi^2}{4611686018427387904}$ \\[3pt]
\hline
&&&&\\[-1em]
37/2 & $\frac{109037877524684653417 \pi^2}{147573952589676412928}$  && 19 & $\frac{96235033003394026579 \pi^2}{147573952589676412928}$ \\[3pt]
\hline
&&&&\\[-1em]
39/2 & $\frac{109807300223973738799 \pi^2}{147573952589676412928}$  && 20 & $\frac{48456823182211506095 \pi^2}{73786976294838206464}$ \\[3pt]
\hline
\end{tabular}
\caption{We give the values of lifts of the state $|\alpha d\rangle^{--}_{++(m,s)}$ for two families where one of the modes has a small fixed mode number and the other mode is varied. We choose to present the first twenty lifts in each of these families.\label{adtab}}%
\end{table}

\subsection{Lifting of $|dd\rangle_{(r,s)}^{\beta B\alpha A}$} \label{ssec.Liftdd}

Lastly, we consider the lift of normalised states of the form
\begin{equation} \label{eq.ddState}
    |dd\rangle_{(r,s)}^{\beta B\alpha A} \equiv \frac{1}{\sqrt{2}} \Big(d^{(1)\beta B}_{-r} d^{(1)\alpha A}_{-s} + d^{(2)\beta B}_{-r} d^{(2)\alpha A}_{-s}\Big) \NSok{1}\NSok{2} \ ,
\end{equation}
where the norm is given by
\begin{align} \label{eq.ddNorm}
    {}_{\ \, (r,s)}^{\delta D\gamma C}\!\langle dd|dd\rangle_{(r,s)}^{\beta B\alpha A} &= \Big(\epsilon^{\alpha\beta} \epsilon^{\gamma\delta} \epsilon_{AB} \epsilon_{CD} \delta_{r+s,1} - \epsilon^{\alpha\gamma} \epsilon^{\beta\delta} \epsilon_{AC}  \epsilon_{BD} \delta_{r,s} \nonumber\\
    &\hspace{4cm}\, + \epsilon^{\alpha\delta} \epsilon^{\beta\gamma} \epsilon_{AD} \epsilon_{BC} H\big[r-\tfrac12\big] \Big) H\big[s-\tfrac12\big] \ ,
\end{align}
with appropriate choices of $SU(2)$ indices as per Appendix~\ref{app.Herm}. The computation of the lift requires the following left-moving amplitude
\begin{equation} \label{eq.Add11def}
    A_{r,s}^{(1)(1)}(w_2,w_1) \equiv \langle\NSo|\big(d^{(1)\delta D}_{s}d^{(1)\gamma C}_{r}\big)\, \Big(G^+_{-,-\frac12}\sigma^-\Big)(w_2) \Big(G^-_{+,-\frac12}\sigma^+\Big)(w_1)\, \big(d^{(1)\beta B,}_{-r}d^{(1)\alpha A}_{-s}\big) |\NSo\rangle \ ,
\end{equation}
where we have chosen all of the modes to act on copy 1. In order to have a non-vanishing state we require
\begin{equation}
    s\geq\frac12\ .
\end{equation}
In mapping to the covering space and removing the spin fields that appear there the initial state transforms as
\begin{align} \label{eq.ddtransi}
    \big(d^{(1)\beta B}_{-r}\! d^{(1)\alpha A}_{-s}\big)_{\!-\infty} &= \oint_{-\infty}\frac{dw_4}{2\pi i}\oint_{-\infty} \frac{dw_3}{2\pi i}\, e^{-r w_4}e^{-s w_3} \, \psi^{\beta B}(w_4) \psi^{\alpha A}(w_3) \nonumber\\
    &= \oint_{0}\frac{dz_4}{2\pi i}\oint_{0} \frac{dz_3}{2\pi i}\, z_4^{-r-\frac12} z_3^{-s-\frac12} \, \psi^{\beta B}(z_4) \psi^{\alpha A}(z_3) \nonumber\\
    &= \oint_{\!-a}\!\frac{dt_4}{2\pi i}\oint_{\!-a}\! \frac{dt_3}{2\pi i} \bigg(\frac{dz_4}{dt_4}\bigg)^{\!\frac12}\!\bigg(\frac{dz_3}{dt_3}\bigg)^{\!\frac12} \, \frac{\psi^{\beta B}(t_4) \psi^{\alpha A}(t_3)}{z_4(t_4)^{r+\frac12} z_3(t_3)^{s+\frac12}} \nonumber\\
    &\longrightarrow \oint_{\!-a}\!\frac{dt_4}{2\pi i}\oint_{\!-a}\! \frac{dt_3}{2\pi i} \bigg(\frac{dz_4}{dt_4}\bigg)^{\!\!\frac12}\!\bigg(\frac{dz_3}{dt_3}\bigg)^{\!\!\frac12}\!\bigg(\frac{t_4-t_1}{t_4-t_2}\bigg)^{\!\!q_\beta}\!\! \bigg(\frac{t_3-t_1}{t_3-t_2}\bigg)^{\!\!q_\alpha}\! \frac{\psi^{\beta B}(t_4) \psi^{\alpha A}(t_3)}{z_4(t_4)^{r+\frac12} z_3(t_3)^{s+\frac12}} \, ,
\end{align}
where the spectral flow transformation was made in the final line and $q_\alpha$ and $q_\beta$ correspond to the values of the $J^3_0$ charge of $d^{\alpha A}_{-s}$ and $d^{\beta B}_{-r}$ respectively. Likewise the final state transforms as
\begin{align} \label{eq.ddtransf}
    \big(d^{(1)\delta D}_{s}d^{(1)\gamma C}_{r}\big)_{\infty} &= \oint_{\infty}\frac{dw_5}{2\pi i}\oint_{\infty} \frac{dw_6}{2\pi i} e^{r w_5} e^{s w_6} \, \psi^{\delta D}(w_6) \psi^{\gamma C}(w_5) \nonumber\\
    &= \oint_{\infty}\frac{dz_5}{2\pi i}\oint_{\infty} \frac{dz_6}{2\pi i} z_5^{r-\frac12} z_6^{s-\frac12} \, \psi^{\delta D}(z_6) \psi^{\gamma C}(z_5) \nonumber\\
    &= \oint_{\infty}\frac{dt_5}{2\pi i}\oint_{\infty} \frac{dt_6}{2\pi i} z_5(t_5)^{r-\frac12} z_6(t_6)^{s-\frac12} \bigg(\frac{dz_6}{dt_6}\bigg)^{\!\frac12}\bigg(\frac{dz_5}{dt_5}\bigg)^{\!\frac12}\, \psi^{\delta D}(t_6) \psi^{\gamma C}(t_5) \nonumber\\
    &\longrightarrow \oint_{\infty}\frac{dt_5}{2\pi i}\oint_{\infty} \frac{dt_6}{2\pi i} \bigg(\frac{dz_6}{dt_6}\bigg)^{\!\!\frac12}\!\bigg(\frac{dz_5}{dt_5}\bigg)^{\!\!\frac12}\!\bigg(\frac{t_6-t_1}{t_6-t_2}\bigg)^{\!\!q_\delta}\! \bigg(\frac{t_5-t_1}{t_5-t_2}\bigg)^{\!\!q_\gamma}\! \frac{\psi^{\delta D}(t_6) \psi^{\gamma C}(t_5)}{z_5(t_5)^{\frac12-r} z_6(t_6)^{\frac12-s}} \, .
\end{align}
This leaves the amplitude \eqref{eq.Add11def} (up to overall factors) as a correlation function of four contours of $\psi$ fields and two insertions of $G$ fields
\begin{align} \label{eq.Add11def2}
    &A_{r,s}^{(1)(1)} \sim \oint_{\infty}\frac{dt_5}{2\pi i} \oint_{\infty} \frac{dt_6}{2\pi i}\oint_{-a}\frac{dt_4}{2\pi i}\oint_{-a} \frac{dt_3}{2\pi i} \frac{z_5(t_5)^{r-\frac12} z_6(t_6)^{s-\frac12}}{z_4(t_4)^{r+\frac12} z_3(t_3)^{s+\frac12}}\bigg(\frac{dz_6}{dt_6}\bigg)^{\!\frac12}\bigg(\frac{dz_5}{dt_5}\bigg)^{\!\frac12} \bigg(\frac{dz_4}{dt_4}\bigg)^{\!\frac12}\bigg(\frac{dz_3}{dt_3}\bigg)^{\!\frac12}\nonumber\\
    &\ \: \times\!\! \bigg(\frac{t_6-t_1}{t_6-t_2}\bigg)^{\!\!q_\delta}\!\!\bigg(\frac{t_5-t_1}{t_5-t_2}\bigg)^{\!\!q_\gamma}\!\!\bigg(\frac{t_4-t_1}{t_4-t_2}\bigg)^{\!\!q_\beta}\!\!\bigg(\frac{t_3-t_1}{t_3-t_2}\bigg)^{\!\!q_\alpha}\! \langle \psi^{\delta D}(t_6) \psi^{\gamma C}(t_5) G^+_-(t_2) G^-_+(t_1) \psi^{\beta B}(t_4) \psi^{\alpha A}(t_3) \rangle \, .
\end{align}
This covering space correlator can be computed straightforwardly by breaking the supercurrents into the free-field bosons and fermions via \eqref{eq.GpsiX} and using the fact that the only bosonic fields in this correlator are from the supercurrents and so they must Wick contract together. The correlator is then given by
\begin{align}
\label{psipsiGGpsipsicorrelator}
    \langle \psi^{\delta D}(t_6)& \psi^{\gamma C}(t_5) G^+_-(t_2) G^-_+(t_1) \psi^{\beta B}(t_4) \psi^{\alpha A}(t_3) \rangle = \frac{-1}{(t_2-t_1)^2}\Bigg[ \nonumber\\
    &\quad-\frac{\epsilon^{CD}\epsilon^{\gamma\delta}}{t_6-t_5}\bigg( \frac{2\epsilon^{AB}\epsilon^{\alpha\beta}}{(t_2-t_1)(t_4-t_3)} - \frac{\epsilon^{\beta+}\epsilon^{\alpha-}\epsilon^{AB}}{(t_2-t_4)(t_1-t_3)} + \frac{\epsilon^{\alpha+}\epsilon^{\beta-}\epsilon^{AB}}{(t_2-t_3)(t_1-t_4)} \bigg) \nonumber\\
    &\quad+\frac{\epsilon^{\delta+}}{t_6-t_2} \bigg( \frac{\epsilon^{AB}\epsilon^{\alpha\beta}\epsilon^{\gamma-}\epsilon^{CD}}{(t_5-t_1)(t_4-t_3)} - \frac{\epsilon^{BC}\epsilon^{\beta\gamma}\epsilon^{\alpha-}\epsilon^{AD}}{(t_5-t_4)(t_1-t_3)} + \frac{\epsilon^{AC}\epsilon^{\alpha\gamma}\epsilon^{\beta-}\epsilon^{BD}}{(t_5-t_3)(t_1-t_4)} \bigg) \nonumber\\
    &\quad+\frac{\epsilon^{\delta-}}{t_6-t_1} \bigg( \frac{\epsilon^{AB}\epsilon^{\alpha\beta}\epsilon^{\gamma+}\epsilon^{CD}}{(t_5-t_2)(t_4-t_3)} - \frac{\epsilon^{BC}\epsilon^{\beta\gamma}\epsilon^{\alpha+}\epsilon^{AD}}{(t_5-t_4)(t_2-t_3)} + \frac{\epsilon^{AC}\epsilon^{\alpha\gamma}\epsilon^{\beta+}\epsilon^{BD}}{(t_5-t_3)(t_2-t_4)} \bigg) \nonumber\\
    &\quad+\frac{\epsilon^{BD}\epsilon^{\beta\delta}}{t_6-t_4}\bigg( \frac{2\epsilon^{AC}\epsilon^{\alpha\gamma}}{(t_2-t_1)(t_5-t_3)} + \frac{\epsilon^{\alpha+}\epsilon^{\gamma-}\epsilon^{AC}}{(t_5-t_1)(t_2-t_3)} + \frac{\epsilon^{\alpha-}\epsilon^{\gamma+}\epsilon^{AC}}{(t_5-t_2)(t_1-t_3)} \bigg) \nonumber\\
    &\quad-\frac{\epsilon^{AD}\epsilon^{\alpha\delta}}{t_6-t_3}\bigg( \frac{2\epsilon^{BC}\epsilon^{\beta\gamma}}{(t_2-t_1)(t_5-t_4)} + \frac{\epsilon^{\beta+}\epsilon^{\gamma-}\epsilon^{BC}}{(t_5-t_1)(t_2-t_4)} + \frac{\epsilon^{\beta-}\epsilon^{\gamma+}\epsilon^{BC}}{(t_5-t_2)(t_1-t_4)} \bigg) \Bigg] \ .
\end{align}
Again, all of the remaining steps in the lifting computation are identical to that described in Section~\ref{ssec.Liftaa}
\begin{equation} \label{eq.ddliftForm}
   E^{(2)}\big(|dd\rangle^{\beta B\alpha A}_{(r,s)}\big) = \frac{\pi\lambda^2}{\big||dd\rangle^{\beta B\alpha A}_{(r,s)}\big|^2} \lim_{T\to\infty} \int^{2\pi}_0 \!d\sigma\, A_{r,s}(w,0) \coth(\tfrac{\bar w}{2}) \ ,
\end{equation}
where $w= \frac{T}{2} + i \sigma$ and $\wb = \frac{T}{2} - i\sigma$ with $T\to \infty$ on this contour and $A_{r,s}(w_2,w_1)$ is the copy-symmetrised amplitude defined analogously to \eqref{eq.Asymm}. For generic $SU(2)$ doublet indices on the $d$ modes in the amplitude \eqref{eq.Add11def} it is not directly related to the second order lift of a particular state of the form \eqref{eq.ddState}. Only when the final state is the Hermitian conjugate of the initial state (as defined in Appendix~\ref{app.Herm}) can we use the relation \eqref{eq.ddliftForm} to obtain a lift.\\
\\
As an example the lifting matrix in mode numbers $r,s$ of the states $|dd\rangle_{(r,s)}^{----}$ is given below (the diagonal states $|dd\rangle^{----}_{s,s}$ vanish due to repeated fermion modes):
\\
\begin{align}
\label{eq.ddmat}
\tikz[remember picture, baseline=(mat.center)]{\node[inner sep=0](mat){$
\begin{pmatrix}
- & \frac{\pi^2}{2} & \frac{\pi^2}{2} & \frac{33 \pi^2}{64} & \frac{17 \pi^2}{32} & \frac{8935 \pi^2}{16384} & \frac{9141 \pi^2}{16384} \\\\
\frac{\pi^2}{2} & - & \frac{41 \pi^2}{64} & \frac{5 \pi^2}{8} & \frac{10383 \pi^2}{16384} & \frac{661 \pi^2}{1024} & \frac{344545 \pi^2}{524288} \\\\
\frac{\pi^2}{2} & \frac{41 \pi^2}{64} & - & \frac{2959 \pi^2}{4096} & \frac{11437 \pi^2}{16384} & \frac{367977 \pi^2}{524288} & \frac{23281 \pi^2}{32768} \\\\
\frac{33 \pi^2}{64} & \frac{5 \pi^2}{8} & \frac{2959 \pi^2}{4096} & - & \frac{409209 \pi^2}{524288} & \frac{24603 \pi^2}{32768} & \frac{50412795 \pi^2}{67108864} \\\\
\frac{17 \pi^2}{32} & \frac{10383 \pi^2}{16384} & \frac{11437 \pi^2}{16384} & \frac{409209 \pi^2}{524288} & - & \frac{221622483 \pi^2}{268435456} & \frac{212700391 \pi^2}{268435456} \\\\
\frac{8935 \pi^2}{16384} & \frac{661 \pi^2}{1024} & \frac{367977 \pi^2}{524288} & \frac{24603 \pi^2}{32768} & \frac{221622483 \pi^2}{268435456} & - & \frac{7408688711 \pi^2}{8589934592} \\\\
\frac{9141 \pi^2}{16384} & \frac{344545 \pi^2}{524288} & \frac{23281 \pi^2}{32768} & \frac{50412795 \pi^2}{67108864} & \frac{212700391 \pi^2}{268435456} & \frac{7408688711 \pi^2}{8589934592} & - \\
\end{pmatrix}
$};}
\begin{tikzpicture}[overlay,remember picture]
\draw[blue,thick,->] node[anchor=south west] (nn1) at (mat.north west)
{increasing $s$} (nn1.east) -- (nn1-|mat.north east) ;
\draw[red,thick,->] node[anchor=north west,align=center, inner xsep=0pt] (nn2) at 
(mat.north east)
{\rotatebox{-90}{increasing $r$}} (nn2.south) -- (nn2.south|-mat.south) 
;
\end{tikzpicture}
\end{align}

\begin{table}[H]
\centering
\begin{tabular}{|c|c|}
\hline
&\\[-1em]
$s$ & $E^{(2)}\big(|dd\rangle_{(3/2,s)}^{----}\big)/\lambda^2$
\\[4pt]
\hline
&\\[-1em]
1/2 & $\frac{\pi^2}{2}$ 
\\[4pt]
\hline
&\\[-1em]
3/2 & -
\\[4pt]
\hline
&\\[-1em]
5/2 & $\frac{41\pi^2}{64}$
\\[4pt]
\hline
&\\[-1em]
7/2 & $\frac{5\pi^2}{8}$ 
\\[4pt]
\hline
&\\[-1em]
9/2 & $\frac{10383\pi^2}{16384}$
\\[4pt]
\hline
&\\[-1em]
11/2 & $\frac{661\pi^2}{1024}$ 
\\[4pt]
\hline
&\\[-1em]
13/2 & $\frac{344545\pi^2}{524288}$ 
\\[4pt]
\hline
&\\[-1em]
15/2 & $\frac{21891\pi^2}{32768}$
\\[4pt]
\hline
&\\[-1em]
17/2 & $\frac{728101955\pi^2}{1073741824}$
\\[4pt]
\hline
&\\[-1em]
19/2 & $\frac{11531551\pi^2}{16777216}$
\\[4pt]
\hline
&\\[-1em]
21/2 & $\frac{23909408127\pi^2}{34359738368}$
\\[4pt]
\hline
&\\[-1em]
23/2 & $\frac{3022575865\pi^2}{4294967296}$ 
\\[4pt]
\hline
&\\[-1em]
25/2 & $\frac{6254815021411\pi^2}{8796093022208}$
\\[4pt]
\hline
&\\[-1em]
27/2 & $\frac{394695881205\pi^2}{549755813888}$
\\[4pt]
\hline
&\\[-1em]
29/2 & $\frac{203894138541041\pi^2}{281474976710656}$
\\[4pt]
\hline
&\\[-1em]
31/2 & $\frac{3212455940071\pi^2}{4398046511104}$
\\[4pt]
\hline
&\\[-1em]
33/2 & $\frac{3394845806727666315\pi^2}{4611686018427387904}$
\\[4pt]
\hline
&\\[-1em]
35/2 & $\frac{13358545555911043\pi^2}{18014398509481984}$ \\[4pt]
\hline
&\\[-1em]
37/2 & $\frac{110190981779977547275\pi^2}{147573952589676412928}$
\\[4pt]
\hline
&\\[-1em]
39/2 & $\frac{13864027231138004091\pi^2}{18446744073709551616}$
\\[4pt]
\hline
\end{tabular}
\caption{We give the first twenty lifts of the states $|dd\rangle^{----}_{(r,s)}$ for the family where one mode has a small fixed mode number ($r=\frac32$) and the other mode is varied. The lifts for the family with $r=1/2$ are equal (up to a shift) to a family of $|\alpha d\rangle$ lifts presented in Section~\ref{ssec.Liftad}; this relation will be discussed in Section~\ref{sec.checks}.\label{ddtab}}
\end{table}

\begin{figure}[H]
    \centering
\fbox{\includegraphics[scale=0.45]{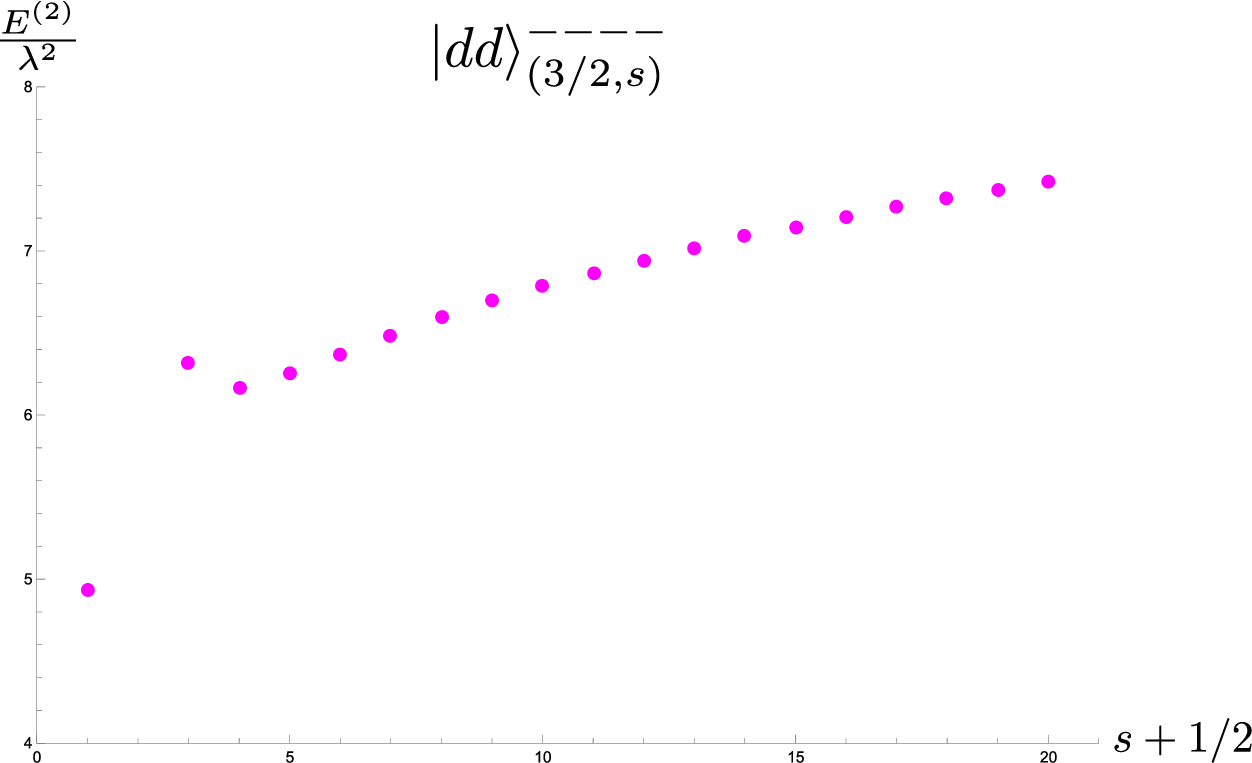}}
    \caption{Plot of the lifts $E^{(2)}\big(|dd\rangle_{(3/2,s)}^{----}\big)/\lambda^2$ for varying $s$. The plot fits to the curve: $-0.41026\; (s+1/2)+3.40665 \sqrt{s+1/2}+0.216748$.}
    \label{fig:8}
\end{figure}

\section{Relations between lifts} \label{sec.checks}

From the low-level lift matrices shown in \eqref{eq.aamat}, \eqref{eq.admat} and \eqref{eq.ddmat} it is clear that there exists some patterns and symmetries in the lifts of these families of two-mode states. In this section we explore some of the relations between lifts of the states considered in Section~\ref{sec.main} that emerge from the action of the $\mathcal{N}=4$ current algebra modes.

\subsection{$L_{-1}$ relations} \label{ssec.Lrelations}

We first consider the relations between lifts due to the action of the stress tensor mode $L_{-1}$ on the initial and final states in the amplitude in \eqref{eq.Xdef}. In what follows, we show that there exist a set of simple relations between the lift of a state and the lift of its $L_{-1}$ descendant involving only a multiplicative factor.

Let us consider a state $|\psi\rangle$ and its $L_{-1}$ descendant $|\psi\rangle$ defined as%
\footnote{One subtlety here is that the descendant state we actually consider is the single-copy descendant $|\phi\rangle^{(1)}\NSok{2} = L_{-1}^{(1)}|\psi\rangle^{(1)}\NSok{2}$, however, in this case this is exactly equal to a global $L_{-1}$ mode acting on $|\psi\rangle\equiv|\psi\rangle^{(1)}\NSok{2}$. This fact is the reason that we consider only $L_{-1}$ modes here.\label{ft1}}
\begin{equation}
    \ket{\phi} \equiv L_{-1}\ket{\psi} \ ,
\end{equation}
where $\ket{\psi}$ has conformal dimension $h_{\psi}$. The stress tensor mode can expressed as a contour integral of the stress tensor field $T(w)$ around the cylinder as
\begin{equation}
    L_{-1} = \frac{1}{2\pi i} \oint dw\, T(w)\, e^{-w} \ ,
\end{equation}
which we will use in the amplitude required in the lift~\eqref{eq.lift} (this was defined as $X(T)$ in \eqref{eq.Xdef}, however, in this section we will use the notation $A(\phi)$)
\begin{align} \label{eq.Lcorr}
    A(\phi) &= \bra{\phi}\int\! d^2w_2\, D(w_2,\bar{w_2})\int\! d^2w_1\, D(w_1,\bar{w_1})\ket{\phi}\nonumber \\
    &= \bra{\psi}L_1\int\! d^2w_2\, D(w_2,\bar{w_2})\int\! d^2w_1\, D(w_1,\bar{w_1})\,L_{-1}\!\ket{\psi} \ .
\end{align}
The general steps in the derivation of these relations between lifts is given pictorially in Figure~\ref{fig:Lcheck}.
\begin{figure}[h]
    \centering
    \includegraphics[width=\textwidth]{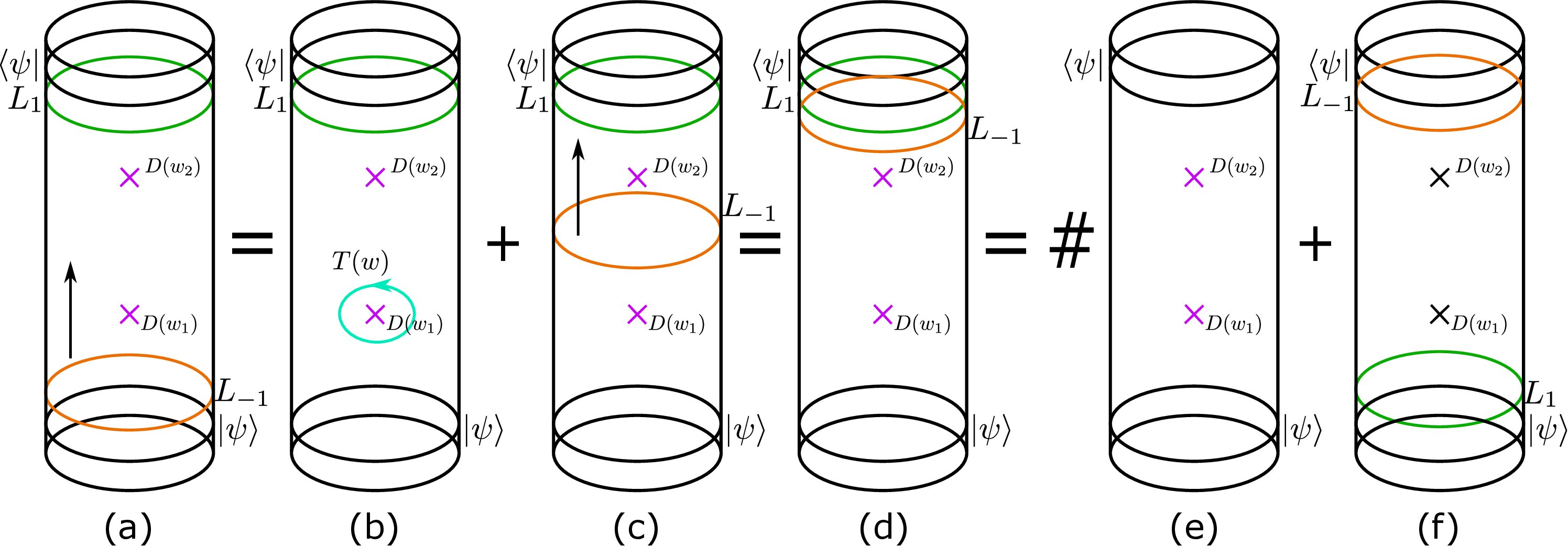}
    \caption{The general tactic in deriving these relations between lifts will be to start with the correlator \eqref{eq.Lcorr} containing descendants of $|\psi\rangle$ as initial and final states (a), deform the contour of the $L_{-1}$ mode through the insertions of the deformation operators (b)-(d) and then commute the $L_{-1}$ mode with the $L_1$ mode from the final state. This leaves (up to constant coefficients) a correlator that computes the lift of the state $|\psi\rangle$ (e) and one that computes the lift of the state $L_1|\psi\rangle$ (f). We have suppressed the trivial right-moving part throughout.}
    \label{fig:Lcheck}
\end{figure}
As shown in Figure~\ref{fig:Lcheck}(a)-(c), the first step is to deform the $L_{-1}$ contour in the amplitude \eqref{eq.Lcorr} which yields a contour encircling the first deformation operator insertion and one wrapped around the cylinder between the two D insertions. We now show that the first of these terms actually vanishes. The contour integral $I_{L}$ of the stress tensor around the insertion of $D(w_1)$ can be evaluated by using the OPE
\begin{equation} \label{eq.TDOPE}
    T(w)D(w_1) \sim \frac{D(w_1)}{(w-w_1)^2}+\frac{\partial_{w_1}D(w_1)}{(w-w_1)}+ \text{reg} \ ,
\end{equation}
yielding
\begin{align}
    I_L &= \frac{1}{2\pi i}\int d^2w_1 \oint_{w_1} dw\, e^{-w}\, T(w)D(w_1) \nonumber \\
    &= \frac{1}{2\pi i}\int d^2w_1 \oint_{w_1} dw\,e^{-w_1}\Big(1 - (w-w_1)+\cdots\Big) T(w)D(w_1)  \nonumber \\
    &= \frac{1}{2\pi i}\int d^2w_1 e^{-w_1}\oint_{w_1} dw \bigg(\frac{D(w_1)}{(w-w_1)^2}+\frac{\partial_{w_1}D(w_1)}{(w-w_1)}-\frac{D(w_1)}{(w-w_1)}\bigg) \nonumber\\
    &= \int d^2w_1 e^{-w_1}\Big(\partial_{w_1}D(w_1) - D(w_1)\Big) = 0 \ ,
\end{align}
where \eqref{eq.TDOPE} was used in the third line and the final integrand is simply a total derivative of $e^{-w_1}D(w_1)$. This argument can be exactly repeated when deforming the $L_{-1}$ contour past the $D(w_2)$ insertion. The correlator \eqref{eq.Lcorr} containing descendant initial and final states is thus equal to the correlator represented by Figure~\ref{fig:Lcheck}(d), \textit{i.e.}
\begin{align} \label{eq.Lcorr2}
    A(\phi) &= \bra{\psi}L_1L_{-1}\int\! d^2w_2\, D(w_2,\bar{w_2})\int\! d^2w_1\, D(w_1,\bar{w_1})\ket{\psi} \nonumber\\
    &= \bra{\psi}\big(2h_{\psi} + L_{-1}L_1\big)\int d^2w_2 D(w_2,\bar{w_2})\int d^2w_1 D(w_1,\bar{w_1})\ket{\psi} \ ,
\end{align}
where the two terms in the last line are displayed in Figure~\ref{fig:Lcheck}(e) and (f). If the state $\psi$ is chosen to be a global conformal primary (that is $L_1|\psi\rangle=0$) then the second term in \eqref{eq.Lcorr2} (depicted in Figure~\ref{fig:Lcheck}(f)) vanishes and we are left with a relation between amplitudes given by
\begin{equation} \label{Lrel}
    A(\phi) = 2h_{\psi} A(\psi) \ .
\end{equation}
Promoting this relation between amplitudes to a precise relation between lifts is discussed in Appendix~\ref{app.B} and we will do this for some explicit cases in Section~\ref{ssec.checks} using the data of Section~\ref{sec.main}.

\subsection{$G^{\alpha}_{\Ad,-\frac12}$ relations} \label{Grelation}

It turns out that very similar relations also exist between a state $|\tilde{\psi}\rangle$ and its $G_{-1/2}$ descendant. Much of the derivation follows along the same lines as in Section~\ref{ssec.Lrelations} with some added complexities from there being multiple choices of $G$ modes. Again, we show that there exist simple relations that link the lifts of different two-mode states of the kind discussed in Section~\ref{sec.main}. One key difference with the relations of Section~\ref{ssec.Lrelations} is that these relate the lifts of states in different families (\textit{i.e.} between the $\alpha\alpha$, $\alpha d$ and $dd$ families of states).

Let us consider a superdescendant of a state $|\tilde{\psi}\rangle$, defined as
\begin{gather} \label{eq.tildephi}
    |\widetilde{\phi}\,\rangle^{\alpha}_{\Ad} \equiv G^\alpha_{\Ad,-\frac12} |\widetilde{\psi}\,\rangle \ ,
\end{gather}
where $|\widetilde{\psi}\,\rangle$ has dimension $h_{\widetilde{\psi}}$ and $J^3_0$ charge $m_{\widetilde{\psi}}$ and the $G$ mode can expressed as a contour integral of a $G$ field as
\begin{equation} \label{eq.Gcontour}
    G^\alpha_{\Ad,-1/2} = \frac{1}{2\pi i} \oint_{C_\tau} dw\, e^{-\frac{1}{2}w}\,G^\alpha_{\Ad}(w) \ ,
\end{equation}
where the contour $C_\tau$ wraps the cylinder at a fixed $\tau$ coordinate. Then the amplitude \eqref{eq.Xdef} we wish to compute for the lift \eqref{eq.lift} is
\begin{align} \label{eq.Gcorr}
    A^{\beta\alpha}_{\Bd\Ad}(\widetilde{\phi}) &= {}^{\ \beta}_{\:\Bd}\langle\widetilde\phi\,|\int\! d^2w_2\, D(w_2,\bar{w}_2)\int\! d^2w_1\, D(w_1,\bar{w}_1)|\widetilde\phi\,\rangle^{\alpha}_{\Ad}\nonumber
    \\
    &= \langle\widetilde\psi\,|G^\beta_{\Bd,1/2}\int\! d^2w_2\, D(w_2,\bar{w}_2)\int\! d^2w_1\, D(w_1,\bar{w}_1)\,G^\alpha_{\Ad,-1/2}|\widetilde\psi\,\rangle \ ,
\end{align}
where it is understood that once a choice of the state \eqref{eq.tildephi} is made, the bra state in the amplitude \eqref{eq.Gcorr} is fixed to be its Hermitian conjugate. For now, however, we keep the indices of the $G$ modes unfixed. Following the same steps as displayed in Figure~\ref{fig:Lcheck} the contour of $G^\alpha_{\Ad}(w)$ (defined in \eqref{eq.Gcontour}) that was acting on the initial state in Figure~\ref{fig:Lcheck}(a) can be deformed through the two $D$ insertions, with the terms containing contour integrals $I^{\alpha}_{\Ad}$ of a $G$ field around a $D$ insertion vanishing (see Figure~\ref{fig:Lcheck}(b) for one of these terms). The first of these terms, around $D(w_1)$, is given by
\begin{align} \label{eq.IG}
    I^{\alpha}_{\Ad} &= \frac1{2\pi i}\int\!d^2w_1 \oint_{C_{w_1}}\!dw \, e^{-\frac{1}{2}w}\,G^\alpha_{\Ad}(w) D(w_1) \nonumber\\
    &= \frac1{2\pi i}\int\!d^2w_1\,e^{-\frac12 w_1} \oint_{C_{w_1}}\!dw \sum_{k=0}^{\infty} \frac{(-1)^k}{2^k k!}(w-w_1)^k\, G^\alpha_{\Ad}(w) D(w_1) \nonumber\\
    &= \int\!d^2w_1\,e^{-\frac12 w_1} \sum_{k=0}^{\infty} \frac{(-1)^k}{2^k k!} \, G^\alpha_{\Ad,k-\frac12} D(w_1) \ ,
\end{align}
where the contour $C_{w_1}$ is centred around $w_1$ and in the last line we used that on such contours the modes on the cylinder are defined as
\begin{equation} \label{eq.Gcontour2}
    G^{\alpha}_{\Ad,s} = \oint_{C_{w_1}}\!\frac{dw}{2\pi i}\, (w-w_1)^{s+\frac12}\, G^{\alpha}_{\Ad} \ .
\end{equation}
As stated in \eqref{eq.deformationDef} the deformation operator can be written in multiple equivalent ways; to analyse \eqref{eq.IG} further it is simplest to choose the representation of $D$ containing a twist operator with $SU(2)_L$ index equal to that of the $G^{\alpha}_{\Ad,-\frac12}$ mode of the starting descendant state \eqref{eq.tildephi}. For instance, for the choice $\alpha=+$ the integral \eqref{eq.IG} becomes
\begin{align} \label{eq.I+G}
    I^{+}_{\Ad} &= \epsilon^{\Cd\Dd}\bar{G}^-_{\dot{D},-\frac12} \int\!d^2w_1\,e^{-\frac12 w_1} \sum_{k=0}^{\infty} \frac{(-1)^k}{2^k k!} \, G^+_{\Ad,k-\frac12} G^-_{\dot{C},-\frac12}\sigma^{++}_2(w_1,\wb_1) \nonumber\\
    &= \epsilon^{\Cd\Dd}\bar{G}^-_{\dot{D},-\frac12} \int\!d^2w_1\,e^{-\frac12 w_1} \sum_{k=0}^{\infty} \frac{(-1)^k}{2^k k!} \, \bigg( \Big\{G^+_{\Ad,k-\frac12}, G^-_{\dot{C},-\frac12}\Big\} + G^-_{\dot{C},-\frac12} G^+_{\Ad,k-\frac12} \bigg)\sigma^{++}_2(w_1,\wb_1) \nonumber\\
    &= -\bar{G}^-_{\dot{A},-\frac12} \int\!d^2w_1\,e^{-\frac12 w_1} \sum_{k=0}^{\infty} \frac{(-1)^k}{2^k k!} \, \bigg( k J^3_{k-1}+L_{k-1} \bigg)\sigma^{++}_2(w_1,\wb_1) \nonumber\\
    &= -\bar{G}^-_{\dot{A},-\frac12} \int\!d^2w_1\,e^{-\frac12 w_1} \, \bigg[ L_{-1} - \frac12 \Big(J^3_{0} + L_0\Big) \bigg] \sigma^{++}_2(w_1,\wb_1) \nonumber\\
    &= -\bar{G}^-_{\dot{A},-\frac12} \int\!d^2w_1\,e^{-\frac12 w_1} \, \bigg( \partial_{w_1} - \frac12 \bigg)\sigma^{++}_2(w_1,\wb_1) = 0 \ ,
\end{align}
where in the third line we use the fact that $\sigma_2^{++}$ is chiral and in the forth line we use that it is killed by positive $J^3$ and $L$ modes. The integral \eqref{eq.I+G} vanishes since the integrand in the final line forms a total derivative. The argument for the case of $I^-_{\Ad}$ follows almost identically using the representation $D(w_1,\wb_1)=\epsilon^{\Cd\Dd} G^+_{\Cd,-\frac12}\bar{G}^+_{\Dd,-\frac12} \sigma_2^{--}(w_1,\wb_1)$ instead. We thus conclude that terms of the form shown in Figure~\ref{fig:Lcheck}(b) with a $G$ contour encircling the insertion of a deformation operator vanish. This leaves us with the amplitude shown in Figure~\ref{fig:Lcheck}(d)
\begin{align} \label{eq.Gcorr2}
    A^{\beta\alpha}_{\Bd\Ad}(\widetilde{\phi}) &= \langle\widetilde{\psi}\,| G^\beta_{\Bd,1/2}G^\alpha_{\Ad,-1/2} \int\! d^2w_2\, D(w_2,\bar{w_2})\int\! d^2w_1\, D(w_1,\bar{w_1})|\widetilde{\psi}\rangle \nonumber\\
    &= \epsilon_{\Bd\Ad} \langle\widetilde{\psi}\,| \Big( \big(\sigma^{aT}\big)^{\beta}_{\gamma}\epsilon^{\gamma\alpha} J^a_0 + \epsilon^{\beta\alpha}L_0 \Big) \int\! d^2w_2\, D(w_2,\bar{w_2})\int\! d^2w_1\, D(w_1,\bar{w_1})|\widetilde{\psi}\rangle \nonumber\\
    &\ \ - \langle\widetilde{\psi}\,| G^\alpha_{\Ad,-1/2}G^\beta_{\Bd,1/2} \int\! d^2w_2\, D(w_2,\bar{w_2})\int\! d^2w_1\, D(w_1,\bar{w_1})|\widetilde{\psi}\rangle \ ,
\end{align}
where the two terms in the last line are those displayed in Figure~\ref{fig:Lcheck}(e) and (f). If the state $\widetilde{\psi}$ is chosen to satisfy $G^{\beta}_{\Bd,\frac12}|\widetilde{\psi}\,\rangle=0$ then the second term in \eqref{eq.Gcorr2} vanishes and we are left with a simple relation between amplitudes. Clearly to relate the amplitude \eqref{eq.Gcorr} to that required in the computation of lifts it is necessary for the quantum numbers of the $G$ mode in the bra state to be opposite to those of the $G$ mode in the ket (an additional overall negative sign may also be required as per the conjugation conventions given in Appendix~\ref{app.Herm}). With this being the case our relations read
\begin{equation} \label{Grel}
    A^{\beta\alpha}_{\Bd\Ad}(\widetilde{\phi}) = K^{\beta\alpha}_{\Bd\Ad} A(\widetilde{\psi}) \ ,
\end{equation}
where
\begin{equation} \label{Kdef}
    K^{\beta\alpha}_{\Bd\Ad} \equiv \epsilon_{\Bd\Ad} \Big( m_{\widetilde{\psi}}\big(\sigma^{3}\big)^{\beta}_{\gamma}\epsilon^{\gamma\alpha} + h_{\widetilde{\psi}}\epsilon^{\beta\alpha} \Big) \ .
\end{equation}
Promoting this relation between amplitudes to a precise relation between lifts is again shown in Appendix~\ref{app.B} and we will do this for some explicit cases in Section~\ref{ssec.checks} using the data of Section~\ref{sec.main}.

\subsection{Using lifting relations to perform checks} \label{ssec.checks}

We will now use the above-derived relations to explain some of the symmetries and properties of the lifting matrices computed in Section~\ref{sec.main}.

\subsubsection{$L_{-1}$ mode on $|\alpha\alpha\rangle_{++++(1,1)}$}

In the lifting matrix of the states $|\alpha\alpha\rangle_{++++(m,n)}$ shown in \eqref{eq.aamat} it is clear that the lifts of the first two states of the top row are both equal to $\frac{\pi^2}{2}$. This equality can be explained by considering the $L_{-1}$ descendant of the level $1$ state
\begin{align}
    L_{-1} |\alpha\alpha\rangle_{++++(1,1)} = 2|\alpha\alpha\rangle_{++++(2,1)} \ .
\end{align}
By using the relation \eqref{eq.LliftRel} we find that the lifts are related by
\begin{align}
     E^{(2)}\big(|\alpha\alpha\rangle_{++++(2,1)}\big) = E^{(2)}\big(|\alpha\alpha\rangle_{++++(1,1)}\big) \ ,
\end{align}
where on the left-hand side we used the fact that
\begin{equation} \label{eq.xphiRel}
    E^{(2)}\big(x |\phi\rangle\big)=E^{(2)}\big(|\phi\rangle\big) \ ,
\end{equation}
for any constant $x$.

\subsubsection{$L_{-1}$ mode on $|dd\rangle_{(3/2,1/2)}^{----}$}

In the lifting matrix of the states $|dd\rangle_{(r,s)}^{----}$ shown in \eqref{eq.ddmat} we see that the lifts in the second and third entries of the first row are equal to $\frac{\pi^2}{2}$. This equality can be explained by considering the $L_{-1}$ descendant of the level $2$ state
\begin{equation}
    L_{-1} |dd\rangle_{(3/2,1/2)}^{----} = |dd\rangle_{(5/2,1/2)}^{----} + |dd\rangle_{(3/2,3/2)}^{----} = |dd\rangle_{(5/2,1/2)}^{----} \ ,
\end{equation}
where the state $|dd\rangle_{(3/2,3/2)}^{----}$ vanishes due to repeated fermion modes. By applying the relation \eqref{eq.LliftRel} we find that
\begin{equation}
    E^{(2)}\big(|dd\rangle_{(5/2,1/2)}^{----}\big) = E^{(2)}\big(|dd\rangle_{(3/2,1/2)}^{----}\big) \ .
\end{equation}
We note that for the relation \eqref{eq.LliftRel} to be applicable here it is necessary that the initial state is annihilated by $L_1$
\begin{equation}
    L_1|dd\rangle^{----}_{(3/2,1/2)} = |dd\rangle^{----}_{(1/2,1/2)} + |dd\rangle^{----}_{(3/2,-1/2)} = 0 \ ,
\end{equation}
due to each state vanishing.

\subsubsection{$G^{-}_{-,\frac12}$ mode on $|\alpha\alpha\rangle_{++++(n,n)}$}

There are also lifts in different families that are equal; by comparing the lifting matrices \eqref{eq.aamat} and \eqref{eq.admat} one sees that the lifts of the diagonal states are equal. We reproduce the relevant shallow lift matrices below: see
\begin{align}
    E^{(2)}\big(|\alpha\alpha\rangle_{++++(m,n)}\big)=\lambda^2
    \begin{pmatrix}
    \tikz[baseline=(char.base)]{\node[draw=red, circle, inner sep=1pt] (char) {$\frac{\pi^2}{2}$};} & \frac{\pi^2}{2} & \frac{9 \pi^2}{16} & \frac{19 \pi^2}{32} & \frac{5045 \pi^2}{8192}\\
    \frac{\pi^2}{2} & \tikz[baseline=(char.base)]{\node[draw=blue, circle, inner sep=1pt] (char) {$\frac{19 \pi^2}{32}$};} & \frac{9 \pi^2}{16} & \frac{2569 \pi^2}{4096} & \frac{10775 \pi^2}{16384} \\
    \frac{9 \pi^2}{16} & \frac{9 \pi^2}{16} & \tikz[baseline=(char.base)]{\node[draw=green, circle, inner sep=1pt] (char) {$\frac{2697 \pi^2}{4096}$};} & \frac{5001 \pi^2}{8192} & \frac{44205 \pi^2}{65536} \\
    \frac{19 \pi^2}{32} & \frac{2569 \pi^2}{4096} & \frac{5001 \pi^2}{8192} & \tikz[baseline=(char.base)]{\node[draw=orange, circle, inner sep=1pt] (char) {$\frac{92751 \pi^2}{131072}$};} & \frac{42555 \pi^2}{65536} \\
    \frac{5045 \pi^2}{8192} & \frac{10775 \pi^2}{16384} & \frac{44205 \pi^2}{65536} & \frac{42555 \pi^2}{65536} & \tikz[baseline=(char.base)]{\node[draw=purple, circle, inner sep=1pt] (char) {$\frac{200592615 \pi^2}{268435456}$};} \\
    \end{pmatrix} \ ,
\end{align}
and
\begin{align}
    E^{(2)}\big(| \alpha d\rangle_{++(n,s)}^{--}\big)=\lambda^2
    \begin{pmatrix}
    \tikz[baseline=(char.base)]{\node[draw=red, circle, inner sep=1pt] (char) {$\frac{\pi^2}{2}$};} & \frac{\pi^2}{2} & \frac{35 \pi^2}{64} & \frac{37 \pi^2}{64} & \frac{9859 \pi^2}{16384} \\
    \frac{\pi^2}{2} & \tikz[baseline=(char.base)]{\node[draw=blue, circle, inner sep=1pt] (char) {$\frac{19 \pi^2}{32}$};} & \frac{19 \pi^2}{32} & \frac{1283 \pi^2}{2048} & \frac{10663 \pi^2}{16384}\\
    \frac{33 \pi^2}{64} & \frac{39 \pi^2}{64} & \tikz[baseline=(char.base)]{\node[draw=green, circle, inner sep=1pt] (char) {$\frac{2697 \pi^2}{4096}$};} & \frac{2697 \pi^2}{4096} & \frac{358269 \pi^2}{524288}\\
    \frac{17 \pi^2}{32} & \frac{2563 \pi^2}{4096} & \frac{5525 \pi^2}{8192} & \tikz[baseline=(char.base)]{\node[draw=orange, circle, inner sep=1pt] (char) {$\frac{92751 \pi^2}{131072}$};} & \frac{92751 \pi^2}{131072}\\
    \frac{8935 \pi^2}{16384} & \frac{10495 \pi^2}{16384} & \frac{361355 \pi^2}{524288} & \frac{378645 \pi^2}{524288} & \tikz[baseline=(char.base)]{\node[draw=purple, circle, inner sep=1pt] (char) {$\frac{200592615 \pi^2}{268435456}$};} \\
    \end{pmatrix}\ .
\end{align} 
This matching of lifts can be explained by considering the superdescendants
\begin{equation} \label{eq.Gaamm}
    G^{-}_{-,\frac12} |\alpha\alpha\rangle_{++++(n,n)} = 2i n| \alpha d\rangle_{++(n,n-1/2)}^{--} \ ,
\end{equation}
and using the relation \eqref{eq.GliftRel} and \eqref{eq.xphiRel} one finds that
\begin{equation}
    E^{(2)}\big(| \alpha d\rangle_{++(n,n-1/2)}^{--}\big) = E^{(2)}\big(|\alpha\alpha\rangle_{++++(n,n)}\big) \ .
\end{equation}
Once again, in order to apply the relation \eqref{eq.GliftRel} it was important that the initial state is annihilated by the current mode in the bra descendant state; \textit{i.e.} in this case that
\begin{equation}
    G^+_{+,-\frac12} |\alpha\alpha\rangle_{++++(n,n)} = 0 \ ,
\end{equation}
which is due to the choice of $SU(2)$ indices.

\subsubsection{Relations between lifts of $|\alpha d\rangle$ states}

Next we look at relationships between the lifts of the $|\alpha d\rangle_{++(m,s)}^{--}$ family of states, as computed in Section~\ref{ssec.Liftad}, for which we display a shallow lifting matrix of $E^{(2)}\big(|\alpha d\rangle_{++(m,s)}^{--}\big)/\lambda^2$:
\begin{align} \label{eq.adliftRels}
    \begin{pmatrix}
    \begin{tikzpicture}[baseline=(current bounding box.center)]
        \node[draw, thick, inner sep=4pt, red] {\tikz[baseline=(char.base)]{\node[draw=black, circle, inner sep=1pt] (char) {$\color{black}\frac{\pi^2}{2}$};}};
    \end{tikzpicture}
    & 
    \begin{tikzpicture}[baseline=(current bounding box.center)]
    \node[draw, thick, inner sep=4pt, red] {$\color{black}\frac{\pi^2}{2}$};
    \end{tikzpicture}
    & \frac{35 \pi^2}{64} & \frac{37 \pi^2}{64} & \frac{9859 \pi^2}{16384} & \frac{10171 \pi^2}{16384} & \frac{333847 \pi^2}{524288} \\\\
    \tikz[baseline=(char.base)]{\node[draw=black, circle, inner sep=1pt] (char) {$\frac{\pi^2}{2}$};}  & 
    \begin{tikzpicture}[baseline=(current bounding box.center)]
        \node[draw, thick, inner sep=4pt, blue] {$\color{black}\frac{19 \pi^2}{32}$};
    \end{tikzpicture}
    & 
    \begin{tikzpicture}[baseline=(current bounding box.center)]
        \node[draw, thick, inner sep=4pt, blue] {$\color{black}\frac{19 \pi^2}{32}$};
    \end{tikzpicture}
    & \frac{1283 \pi^2}{2048} & \frac{10663 \pi^2}{16384} & \frac{175711 \pi^2}{262144} & \frac{179969 \pi^2}{262144} \\\\
    \frac{33 \pi^2}{64} & \frac{39 \pi^2}{64} & 
    \begin{tikzpicture}[baseline=(current bounding box.center)]
        \node[draw, thick, inner sep=4pt, green] {$\color{black}\frac{2697 \pi^2}{4096}$};
    \end{tikzpicture}
    & 
    \begin{tikzpicture}[baseline=(current bounding box.center)]
        \node[draw, thick, inner sep=4pt, green] {$\color{black}\frac{2697 \pi^2}{4096}$};
    \end{tikzpicture}
    & \frac{358269 \pi^2}{524288} & \frac{368643 \pi^2}{524288} & \frac{48290217 \pi^2}{67108864} \\\\
    \frac{17 \pi^2}{32} & \frac{2563 \pi^2}{4096} & \frac{5525 \pi^2}{8192} & 
    \begin{tikzpicture}[baseline=(current bounding box.center)]
        \node[draw, thick, inner sep=4pt, orange] {$\color{black}\frac{92751 \pi^2}{131072}$};
    \end{tikzpicture}
    & 
    \begin{tikzpicture}[baseline=(current bounding box.center)]
        \node[draw, thick, inner sep=4pt, orange] {$\color{black}\frac{92751 \pi^2}{131072}$};
    \end{tikzpicture}
    & \frac{48833787 \pi^2}{67108864} & \frac{99898097 \pi^2}{134217728} \\\\
    \frac{8935 \pi^2}{16384} & \frac{10495 \pi^2}{16384} & \frac{361355 \pi^2}{524288} & \frac{378645 \pi^2}{524288} & 
    \begin{tikzpicture}[baseline=(current bounding box.center)]
        \node[draw, thick, inner sep=4pt, purple] {$\color{black}\frac{200592615 \pi^2}{268435456}$};
    \end{tikzpicture}
    & 
    \begin{tikzpicture}[baseline=(current bounding box.center)]
        \node[draw, thick, inner sep=4pt, purple] {$\color{black}\frac{200592615 \pi^2}{268435456}$};
    \end{tikzpicture}
    & \frac{6562949705 \pi^2}{8589934592} \\\\
    \frac{9141 \pi^2}{16384} & \frac{171381 \pi^2}{262144} & \frac{184155 \pi^2}{262144} & \frac{24675753 \pi^2}{33554432} & \frac{204097593 \pi^2}{268435456} & 
    \begin{tikzpicture}[baseline=(current bounding box.center)]
        \node[draw, thick, inner sep=4pt, cyan] {$\color{black}\frac{3351953103 \pi^2}{4294967296}$};
    \end{tikzpicture}
    & 
    \begin{tikzpicture}[baseline=(current bounding box.center)]
        \node[draw, thick, inner sep=4pt, cyan] {$\color{black}\frac{3351953103 \pi^2}{4294967296}$};
    \end{tikzpicture} \\
    \end{pmatrix}
\end{align}
This lifting matrix displays some interesting patterns that we will now explore. Firstly there is the equality of lifts on the diagonal with those on the first off diagonal on the upper-half triangle. This can be understood by considering the application of two $G$ modes to a diagonal term (for which $s=m-\frac12$) since
\begin{equation} \label{eq.GGad}
    G^{-}_{-,-1/2} G^{+}_{+,-1/2} |\alpha d\rangle_{++(m,m-1/2)}^{--} = i\,G^{-}_{-,-1/2} | \alpha \alpha\rangle_{++++(m,m)} = 2im| \alpha d\rangle_{++(m,m+1/2)}^{--} \ .
\end{equation}
We can then use the relation \eqref{eq.GliftRel} at each step of \eqref{eq.GGad} to obtain the simple equality
\begin{equation} \label{eq.EGGad}
    E^{(2)}\big(|\alpha d\rangle_{++(m,m-1/2)}^{--}\big) = E^{(2)}\big(| \alpha d\rangle_{++(m,m+1/2)}^{--}\big) \ .
\end{equation}
The use of these two relations is justified due to the fact that $G^-_{-,\frac12}|\alpha d\rangle_{++(m,m-1/2)}^{--}=0$ and $G^+_{+,\frac12}| \alpha \alpha\rangle_{++++(m,m)}=0$ and that at each step in \eqref{eq.GGad} only one state is created and hence \eqref{eq.GliftRel} could be straightforwardly applied.

Secondly, there is also a single pair of equal lifts in the first column of the lifting matrix \eqref{eq.adliftRels} (in black circles) which can be understood from the action of $L_{-1}$ on the level-$3/2$ state in this family
\begin{equation} \label{eq.Lad}
    L_{-1}| \alpha d\rangle_{++(1,1/2)}^{--}= |\alpha d\rangle_{++(2,1/2)}^{--} + |\alpha d\rangle_{++(1,3/2)}^{--} \ .
\end{equation}
The $L_{-1}$ relation between lifts \eqref{eq.LliftRel} can then be applied, since the condition $L_1| \alpha d\rangle_{++(1,1/2)}^{--}=0$ holds, yielding
\begin{equation} \label{eq.ELad1}
    E^{(2)}\big(L_{-1} |\alpha d\rangle_{++(1,1/2)}^{--}\big) = E^{(2)}\big(|\alpha d\rangle_{++(1,1/2)}^{--}\big) \ .
\end{equation}
Due to the descendant state \eqref{eq.Lad} being a sum of two basis states the lifting relation \eqref{eq.ELad1} cannot be immediately extrapolated to a relation between elements of the lifting matrix \eqref{eq.adliftRels}. Using the identity \eqref{eq.Ephi12} with $|\phi_1\rangle = |\alpha d\rangle_{++(2,1/2)}^{--}$ and $|\phi_2\rangle = |\alpha d\rangle_{++(1,3/2)}^{--}$ along with the fact that $E^{(2)}(\phi_1;\phi_2)=0$ and $\langle\phi_1|\phi_2\rangle=0$ here, \eqref{eq.GliftRel} then gives
\begin{align} \label{eq.Ead2}
    E^{(2)}\big(|\alpha d\rangle_{++(2,1/2)}^{--}\big) = \frac12\bigg[ 3E^{(2)}\big(|\alpha d\rangle_{++(1,1/2)}^{--}\big) - E^{(2)}\big(|\alpha d\rangle_{++(1,3/2)}^{--}\big) \bigg] \ .
\end{align}
Using the previously derived relation \eqref{eq.EGGad} with $m=1$ we then obtain
\begin{equation} \label{eq.ELad2}
    E^{(2)}\big(|\alpha d\rangle_{++(2,1/2)}^{--}\big) = E^{(2)}\big(|\alpha d\rangle_{++(1,1/2)}^{--}\big) \ ,
\end{equation}
as seen from the lifting matrix \eqref{eq.adliftRels}.

\subsection{Vanishing lift of the stress tensor from free field realisation}

As a final check that encompasses a large number of the lifts from Sections~\ref{ssec.Liftaa} and \ref{ssec.Liftdd} we compute the lift of the stress tensor. Clearly, from both general principles and from the fact that any single-mode single-copy excitation of the global NS vacuum can be written as a single global mode on the vacuum, this state has vanishing lift. However, from the perspective of the free-field realisation of the CFT the stress tensor state is given in terms of bosons and fermions as \eqref{eq.Lfree} and its modes can be written as
\begin{align}
    L_m &= -\frac{1}{2} \sum_n \epsilon^{BA}\epsilon^{\dot{B}\dot{A}}\alpha_{B\dot{B},n}\alpha_{A\dot{A},m-n} -\frac{1}{2}\sum_r \bigg(m-r+\frac{1}{2}\bigg)\epsilon_{\beta\alpha}\epsilon_{BA} d^{\beta B}_r d^{\alpha A}_{m-r} \ .
\end{align}
The stress tensor initial state is then given by
\begin{align} \label{eq.L-2def}
    L_{-2}|\NSo\rangle &= -\frac{1}{2} \sum_n \epsilon^{BA}\epsilon^{\dot{B}\dot{A}} |\alpha\alpha\rangle_{B\Bd A\Ad(-n,n+2)} -\frac{1}{2}\sum_r \bigg(\!-2-r+\frac{1}{2}\bigg) \epsilon_{\beta\alpha}\epsilon_{BA} |dd\rangle^{\beta B\alpha A}_{(-r,r+2)} \nonumber\\
    &= -\frac{1}{2} \epsilon^{BA}\epsilon^{\dot{B}\dot{A}} |\alpha\alpha\rangle_{B\Bd A\Ad(1,1)} + \frac{1}{2}\epsilon_{\beta\alpha}\epsilon_{BA} |dd\rangle^{\beta B\alpha A}_{(1/2,3/2)} \ ,
\end{align}
and so the lift of $L_{-2}|\NSo\rangle$ in terms of the states of Sections~\ref{ssec.Liftaa} and \ref{ssec.Liftdd} is non-trivial and its vanishing must come about from cancellation between terms.

In order to compute the lift of $L_{-2}|\NSo\rangle$ from \eqref{eq.lift} it is necessary to compute the lift-moving amplitude
\begin{equation} \label{eq.A11L}
    A^{(1)(1)}_{L_{-2}}(w_2,w_1) \equiv \langle\NSo|\big(L_2^{(1)}\big)\, \Big(G^+_{-,-\frac12}\sigma^-\Big)(w_2) \Big(G^-_{+,-\frac12}\sigma^+\Big)(w_1)\, \big(L_{-2}^{(1)}\big) |\NSo\rangle \ ,
\end{equation}
with the initial and final states broken down as \eqref{eq.L-2def}. This amplitude has the initial and final excitations placed on copy $1$; as in the computations of Section~\ref{sec.main}, the amplitude will then be symmetrised over copies. In this amplitude there will then be four types of contributions, where the initial and final states can each either be of the form $|\alpha\alpha\rangle$ or $|dd\rangle$. The contributions with both initial and final states being of the same family will be related to the lifts computed in Sections~\ref{ssec.Liftaa} and \ref{ssec.Liftdd}, whereas the contributions with initial and final states being from different families will require additional computation.

\subsubsection{$\alpha\alpha-\alpha\alpha$ contributions}

The contributions to \eqref{eq.A11L} with initial and final states being of the $|\alpha\alpha\rangle$ family are given by
\begin{align}
    A^{(1)(1)}_{\alpha\alpha\alpha\alpha} &\equiv \frac{1}{4}\epsilon^{DC}\epsilon^{\dot{D}\dot{C}}\epsilon^{BA}\epsilon^{\dot{B}\dot{A}} \nonumber\\
    &\quad\times\langle\NSo|\big(\alpha^{(1)}_{D\Dd,1}\alpha^{(1)}_{C\Cd,1}\big)\, \Big(G^+_{-,-\frac12}\sigma^-\Big)(w_2) \Big(G^-_{+,-\frac12}\sigma^+\Big)(w_1)\, \big(\alpha^{(1)}_{B\Bd,-1}\alpha^{(1)}_{A\Ad,-1}\big) |\NSo\rangle \ .
\end{align}
The terms in the sums over $SU(2)$ indices where the quantum numbers of the final-state modes are opposite to those of the initial-state modes are directly related to the lifts of those states, however, all of the terms are computed exactly as in Section~\ref{ssec.Liftaa}, requiring a computation on the $t$-plane of the form
\begin{align} \label{L-2ampaaaa}
    A^{(1)(1)}_{\alpha\alpha\alpha\alpha} &\sim \oint_{\infty}\frac{dt_5}{2\pi i} \oint_{\infty}\frac{dt_6}{2\pi i}\oint_{-a}\frac{dt_4}{2\pi i}\oint_{-a} \frac{dt_3}{2\pi i} \frac{z_5(t_5) z_6(t_6)}{z_4(t_4) z_3(t_3)} \epsilon^{DC}\epsilon^{\dot{D}\dot{C}}\epsilon^{BA}\epsilon^{\dot{B}\dot{A}} \nonumber\\
    &\qquad\qquad\qquad\qquad\times\langle \partial X_{D\Dd}(t_6) \partial X_{C\Cd}(t_5) G^+_-(t_2) G^-_+(t_1) \partial X_{B\dot{B}}(t_4) \partial X_{A\dot{A}}(t_3) \rangle \ .
\end{align}
The correlator of fields on the $t$-plane in \eqref{L-2ampaaaa} is given in \eqref{XXGGXXcorrelator}. Performing the integrals over the insertions of the deformation operators and symmetrising over copies yields a total $\alpha\alpha-\alpha\alpha$ contribution to the lift of $L_{-2}|\NSo\rangle$ of
\begin{equation} \label{eq.Iaaaa}
    E^{(2)}_{\alpha\alpha\alpha\alpha}\big(L_{-2}|\NSo\rangle\big) = \frac{3\pi^2}{4}\lambda^2 \ .
\end{equation}

\subsubsection{$\alpha\alpha-dd$ contributions}

The contributions to \eqref{eq.A11L} with initial states being of the $|dd\rangle$ family and final states of the $|\alpha\alpha\rangle$ family are given by
\begin{align}
    A^{(1)(1)}_{\alpha\alpha dd} &\equiv \frac{1}{4}\epsilon^{DC}\epsilon^{\dot{D}\dot{C}}\epsilon_{\beta\alpha}\epsilon_{BA} \nonumber\\
    &\qquad\times\langle\NSo|\big(\alpha^{(1)}_{D\Dd,1}\alpha^{(1)}_{C\Cd,1}\big)\, \Big(G^+_{-,-\frac12}\sigma^-\Big)(w_2) \Big(G^-_{+,-\frac12}\sigma^+\Big)(w_1)\, \big(d^{(1)\beta B}_{-1/2} d^{(1)\alpha A}_{-3/2}\big) |\NSo\rangle \ .
\end{align}
None of these terms are directly related to lifts, however, the computation of these contributions requires only a small modification to methods of Section~\ref{sec.main} to yield (on the $t$-plane)
\begin{align} \label{L-2ampaadd}
    A^{(1)(1)}_{\alpha\alpha dd} &\sim \oint_{\infty}\frac{dt_5}{2\pi i} \oint_{\infty}\frac{dt_6}{2\pi i}\oint_{-a}\frac{dt_4}{2\pi i}\oint_{-a} \frac{dt_3}{2\pi i} \frac{z_5(t_5) z_6(t_6)}{z_4(t_4) z_3(t_3)^2} \bigg(\frac{dz_4}{dt_4}\bigg)^{\!\frac12}\!\bigg(\frac{dz_3}{dt_3}\bigg)^{\!\frac12}\!\bigg(\frac{t_4-t_1}{t_4-t_2}\bigg)^{\!q_\beta}\!\bigg(\frac{t_3-t_1}{t_3-t_2}\bigg)^{\!q_\alpha} \nonumber\\
    &\qquad\qquad\times\epsilon^{DC}\epsilon^{\dot{D}\dot{C}}\epsilon_{\beta\alpha}\epsilon_{BA}\langle \partial X_{D\Dd}(t_6) \partial X_{C\Cd}(t_5) G^+_-(t_2) G^-_+(t_1) \psi^{\beta B}(t_4) \psi^{\alpha A}(t_3) \rangle \  .
\end{align}
The $t$-plane correlator above can be computed by expanding the $G$ fields via \eqref{eq.Gfree} and using Wick contractions
\begin{align} \label{eq.XXGGPP}
    \langle &\partial X_{D\Dd}(t_6) \partial X_{C\Cd}(t_5) G^+_-(t_2) G^-_+(t_1) \psi^{\beta B}(t_4) \psi^{\alpha A}(t_3) \rangle = \frac{-\epsilon_{CD}\epsilon_{\Cd\Dd}}{(t_6-t_5)^2(t_2-t_1)^2}\Bigg[ \frac{2\epsilon^{\alpha\beta}\epsilon^{AB}}{(t_4-t_3)(t_2-t_1)} \nonumber\\
    &- \frac{\epsilon^{+\beta}\epsilon^{-\alpha}\epsilon^{AB}}{(t_2-t_4)(t_1-t_3)} - \frac{\epsilon^{-\beta}\epsilon^{+\alpha}\epsilon^{AB}}{(t_2-t_3)(t_1-t_4)}\Bigg] + \frac{\epsilon_{\Dd-}\epsilon_{\Cd+}}{(t_6-t_2)^2(t_5-t_1)^2} \Bigg[ \frac{-\epsilon^{\alpha\beta}\epsilon^{AB}\epsilon_{CD}}{(t_4-t_3)(t_2-t_1)} \nonumber\\
    &- \frac{\epsilon^{+\beta}\epsilon^{-\alpha}\delta^B_D\delta^A_C}{(t_2-t_4)(t_1-t_3)} + \frac{\epsilon^{-\beta}\epsilon^{+\alpha}\delta^A_D\delta^B_C}{(t_2-t_3)(t_1-t_4)} \Bigg] + \frac{\epsilon_{\Dd+}\epsilon_{\Cd-}}{(t_6-t_1)^2(t_5-t_2)^2} \Bigg[ \frac{\epsilon^{\alpha\beta}\epsilon^{AB}\epsilon_{CD}}{(t_4-t_3)(t_2-t_1)} \nonumber\\
    &- \frac{\epsilon^{+\beta}\epsilon^{-\alpha}\delta^A_D\delta^B_C}{(t_2-t_4)(t_1-t_3)} + \frac{\epsilon^{-\beta}\epsilon^{+\alpha}\delta^B_D\delta^A_C}{(t_2-t_3)(t_1-t_4)} \Bigg] \ .
\end{align}
Performing the integrals over the insertions of the deformation operators and symmetrising over copies yields a total $\alpha\alpha-dd$ contribution to the lift of $L_{-2}|\NSo\rangle$ of
\begin{equation} \label{eq.Iaadd}
    E^{(2)}_{\alpha\alpha dd}\big(L_{-2}|\NSo\rangle\big) = -\frac{3\pi^2}{4}\lambda^2 \ .
\end{equation}

\subsubsection{$dd-\alpha\alpha$ contributions}

The contributions to \eqref{eq.A11L} with initial states being of the $|\alpha\alpha\rangle$ family and final states of the $|dd\rangle$ family are given by
\begin{align}
    A^{(1)(1)}_{dd\alpha\alpha} &\equiv \frac{1}{4}\epsilon_{\delta\gamma}\epsilon_{DC}\epsilon^{BA}\epsilon^{\dot{B}\dot{A}} \nonumber\\
    &\qquad\times\langle\NSo|\big(d^{(1)\delta D}_{3/2} d^{(1)\gamma C}_{1/2}\big)\, \Big(G^+_{-,-\frac12}\sigma^-\Big)(w_2) \Big(G^-_{+,-\frac12}\sigma^+\Big)(w_1)\, \big(\alpha^{(1)}_{B\Bd,-1}\alpha^{(1)}_{A\Ad,-1}\big) |\NSo\rangle \ .
\end{align}
None of these terms are directly related to lifts, however, the computation of these contributions requires only a small modification to methods of Section~\ref{sec.main} to yield (on the $t$-plane)
\begin{align} \label{L-2ampddaa}
    A^{(1)(1)}_{dd \alpha\alpha } &\sim \oint_{\infty}\frac{dt_5}{2\pi i} \oint_{\infty}\frac{dt_6}{2\pi i}\oint_{-a}\frac{dt_4}{2\pi i}\oint_{-a} \frac{dt_3}{2\pi i} \frac{ z_6(t_6)}{z_4(t_4) z_3(t_3)} \bigg(\frac{dz_6}{dt_6}\bigg)^{\!\frac12}\!\bigg(\frac{dz_5}{dt_5}\bigg)^{\!\frac12}\!\bigg(\frac{t_6-t_1}{t_6-t_2}\bigg)^{\!q_\delta}\!\bigg(\frac{t_5-t_1}{t_5-t_2}\bigg)^{\!q_\gamma} \nonumber\\
    &\qquad\qquad\times\epsilon_{\delta\gamma}\epsilon_{DC}\epsilon^{BA}\epsilon^{\dot{B}\dot{A}}\langle \psi^{\delta D}(t_6) \psi^{\gamma C}(t_5) G^+_-(t_2) G^-_+(t_1) \partial X_{B\dot{B}}(t_4) \partial X_{A\dot{A}}(t_3) \rangle \ .
\end{align}
The $t$-plane correlator above can be computed by expanding the $G$ fields via \eqref{eq.Gfree} and using Wick contractions
\begin{align} \label{eq.PPGGXX}
    \langle &\psi^{\delta D}(t_6) \psi^{\gamma C}(t_5) G^+_-(t_2) G^-_+(t_1) \partial X_{B\dot{B}}(t_4) \partial X_{A\dot{A}}(t_3) \rangle = \frac{-\epsilon_{CD}\epsilon_{\Cd\Dd}}{(t_4-t_3)^2(t_2-t_1)^2}\Bigg[ \frac{2\epsilon^{\alpha\beta}\epsilon^{AB}}{(t_6-t_5)(t_2-t_1)} \nonumber\\
    &+ \frac{\epsilon^{\beta-}\epsilon^{\alpha+}\epsilon^{AB}}{(t_6-t_2)(t_5-t_1)} + \frac{\epsilon^{\beta+}\epsilon^{\alpha-}\epsilon^{AB}}{(t_6-t_1)(t_5-t_2)}\Bigg] + \frac{\epsilon_{+\Dd}\epsilon_{-\Cd}}{(t_2-t_4)^2(t_1-t_3)^2} \Bigg[ \frac{\epsilon^{\alpha\beta}\epsilon^{AB}\epsilon_{CD}}{(t_6-t_5)(t_2-t_1)} \nonumber\\
    &- \frac{\epsilon^{\beta-}\epsilon^{\alpha+}\delta^B_D\delta^A_C}{(t_6-t_2)(t_5-t_1)} + \frac{\epsilon^{\beta+}\epsilon^{\alpha-}\delta^A_D\delta^B_C}{(t_6-t_1)(t_5-t_2)} \Bigg] + \frac{\epsilon_{-\Dd}\epsilon_{+\Cd}}{(t_2-t_3)^2(t_1-t_4)^2} \Bigg[ \frac{-\epsilon^{\alpha\beta}\epsilon^{AB}\epsilon_{CD}}{(t_6-t_5)(t_2-t_1)} \nonumber\\
    &- \frac{\epsilon^{\beta-}\epsilon^{\alpha+}\delta^A_D\delta^B_C}{(t_6-t_2)(t_5-t_1)} + \frac{\epsilon^{\beta+}\epsilon^{\alpha-}\delta^B_D\delta^A_C}{(t_6-t_1)(t_5-t_2)} \Bigg] \ .
\end{align}
Performing the integrals over the insertions of the deformation operators and symmetrising over copies yields a total $dd-\alpha\alpha$ contribution to the lift of $L_{-2}|\NSo\rangle$ of
\begin{equation} \label{eq.Iddaa}
    E^{(2)}_{dd \alpha\alpha}\big(L_{-2}|\NSo\rangle\big) = -\frac{3\pi^2}{4}\lambda^2 \ .
\end{equation}

\subsubsection{$dd-dd$ contributions}

The contributions to \eqref{eq.A11L} with initial and final states being of the $|dd\rangle$ family are given by
\begin{align}
    A^{(1)(1)}_{dd dd} &\equiv \frac{1}{4}\epsilon_{\delta\gamma}\epsilon_{DC}\epsilon_{\beta\alpha}\epsilon_{BA} \nonumber\\
    &\qquad\times\langle\NSo|\big(d^{(1)\delta D}_{3/2} d^{(1)\gamma C}_{1/2}\big)\, \Big(G^+_{-,-\frac12}\sigma^-\Big)(w_2) \Big(G^-_{+,-\frac12}\sigma^+\Big)(w_1)\, \big(d^{(1)\beta B}_{-1/2} d^{(1)\alpha A}_{-3/2}\big) |\NSo\rangle \ .
\end{align}
The terms in the sums over $SU(2)$ indices where the quantum numbers of the final-state modes are opposite to those of the initial-state modes are directly related to the lifts of those states, however, all of the terms are computed exactly as in Section~\ref{ssec.Liftaa}, requiring a computation on the $t$-plane of the form
\begin{align} \label{L-2ampdddd}
    A^{(1)(1)}_{dd dd} &\sim \oint_{\infty}\!\frac{dt_5}{2\pi i} \oint_{\infty}\! \frac{dt_6}{2\pi i}\oint_{-a}\!\frac{dt_4}{2\pi i}\oint_{-a}\! \frac{dt_3}{2\pi i} \frac{ z_6(t_6)}{z_4(t_4) z_3(t_3)^{2}} \bigg(\frac{dz_6}{dt_6}\frac{dz_5}{dt_5}\frac{dz_4}{dt_4}\frac{dz_3}{dt_3}\bigg)^{\!\!\frac12} \!\bigg(\frac{t_6-t_1}{t_6-t_2}\bigg)^{\!\!q_\delta}\bigg(\frac{t_5-t_1}{t_5-t_2}\bigg)^{\!\!q_\gamma} \nonumber\\
    & \times\!\bigg(\frac{t_4-t_1}{t_4-t_2}\bigg)^{\!\!q_\beta}\!\bigg(\frac{t_3-t_1}{t_3-t_2}\bigg)^{\!\!q_\alpha}\epsilon_{\delta\gamma}\epsilon_{\beta\alpha} \epsilon_{DC} \epsilon_{BA} \langle \psi^{\delta D}(t_6) \psi^{\gamma C}(t_5) G^+_-(t_2) G^-_+(t_1) \psi^{\beta B}(t_4) \psi^{\alpha A}(t_3) \rangle \ .
\end{align}
The correlator of fields on the $t$-plane in \eqref{L-2ampaaaa} is given in \eqref{psipsiGGpsipsicorrelator}. Performing the integrals over the insertions of the deformation operators and symmetrising over copies yields a total $dd-dd$ contribution to the lift of $L_{-2}|\NSo\rangle$ of
\begin{equation} \label{eq.Idddd}
    E^{(2)}_{dddd}\big(L_{-2}|\NSo\rangle\big) = \frac{3\pi^2}{4}\lambda^2 \ .
\end{equation}
The total lift of the state $L_{-2}|\NSo\rangle$ is then obtained by summing the four types of contributions from \eqref{eq.Iaaaa}, \eqref{eq.Iaadd}, \eqref{eq.Iddaa} and  \eqref{eq.Idddd} to get
\begin{align}
    E^{(2)}\big(L_{-2}|\NSo\rangle\big) &= E^{(2)}_{\alpha\alpha\alpha\alpha}\big(L_{-2}|\NSo\rangle\big) + E^{(2)}_{\alpha\alpha dd}\big(L_{-2}|\NSo\rangle\big) + E^{(2)}_{dd\alpha\alpha}\big(L_{-2}|\NSo\rangle\big) + E^{(2)}_{dddd}\big(L_{-2}|\NSo\rangle\big) \nonumber\\
    &= 0 \ ,
\end{align}
as expected. This provides a non-trivial check the low-level results of Sections~\ref{ssec.Liftaa} and \ref{ssec.Liftdd}.

\section{Discussion} \label{sec.conc}

The D1-D5 CFT has a set of BPS states that are simply described at the orbifold point: all states with purely left moving (or purely right moving) excitations are BPS at this point in moduli space. As we move away from this point some states lift, with the number of unlifted states being bounded from below by an index. However, the question of which states lift and by how much remains. This pattern of lifting is an interesting issue to study and a full understanding of this problem has proved to be elusive so far. Progress has been made in studying the lift of various families of states; some of these computations were summarised in the introduction. The present paper studied a new family of states: those with two oscillator excitations on a single copy of the $c=6$ seed CFT.

The expectation value of the lift for a general state in this family was obtained in terms of a fixed number of nested contour integrals on a given integrand; this integrand depends on the mode numbers of the oscillators in the state. We evaluate these contour integrals to obtain the explicit value of the lift for various subfamilies of states. Explicit lifting matrices for mode numbers up to order $7$ were presented for examples of each of the three types of states in \eqref{eq.aamat}, \eqref{eq.admat} and \eqref{eq.ddmat}. Explicit values of lifts were also computed for two types of subfamilies: (i) the states where one mode number was held to be small and the other was allowed to grow large (we presented the first $20$ lifts) and (ii) where both mode numbers were taken to have the same value; this mode number reached up to a value $20$. This explicit data can be found in tables~\ref{aatab}, \ref{adtab} and \ref{ddtab}. The choice to present explicit lifts for these states was purely for space and computational reasons; the method described in this paper can scale arbitrarily given the resources.

We note that the lifts found in this paper are lifts in the theory for a general number $N$ of copies of the $c=6$ theory. The lifting method used sees only two copies of the seed CFT at a time and thus it is sufficient to consider only two copies in intermediate steps. As given in \eqref{eq.GeneralN}, the fact that there are $N$ copies in total can be easily reinstated using combinatorics. Also of note is that one could equally use a different basis of 2-mode states from what we have used; one where states are grouped into representations of the various $SU(2)$ symmetries of the algebra. In this basis the fact that the lifts of states in the same multiplet are equal would be explicit. However, we find it more convenient to use the basis of Section~\ref{sec.main} with this method of computing lifts. In either case, the additional relations between lifts discussed in Section~\ref{sec.checks} that stem from superconformal Ward identities would still be present.
Having explicit values of the lift up to such high levels allows one to observe a smooth curve passing through the plot of lifts (see e.g. Figure~\ref{fig:aa1}). In \cite{Guo:2022ifr} it was found that for single-copy superconformal primary states of high dimension $h$, the lift grew as $\sim \sqrt{h}$. In the present case, we do not have a closed form expression for the lift for all levels, but the lifts we have explicitly computed do appear to follow a behavior $\sim \sqrt{h}$ for large $h$. It will be interesting to find out if this $\sim\sqrt{h}$ behavior is a general feature of lifts at large dimensions.

\section*{Acknowledgements}

We would like to thank Bin Guo for discussions. This work is supported in part by DOE grant DE-SC0011726.

\appendix

\section{The $\mathcal N=4$ superconformal algebra} \label{app_cft}

We follow the notation of appendix A of \cite{Hampton:2018ygz}. As described in Section~{\ref{ssec.symmCFT}} the indices $\alpha=(+,-)$ and $\bar \alpha=(+,-)$ correspond to the subgroups $SU(2)_L$ and $SU(2)_R$ arising from rotations on $S^3$ and the indices $A=(+,-)$ and $\dot A=(+,-)$ correspond to the subgroups $SU(2)_1$ and $SU(2)_2$ arising from rotations in $T^4$. We use the convention
\be
    \epsilon_{+-}=1 \quad,\quad\epsilon^{+-}=-1 \ .
\ee

\subsection{Commutation relations} \label{app.comms}

The commutation relations for the small $\mathcal N=4$ superconformal algebra are
\begin{subequations} \label{app com currents}
    \begin{align}
        \big[L_m,L_n\big] &= \frac{c}{12}m(m^2-1)\delta_{m+n,0}+ (m-n)L_{m+n} \ , \label{LLcomm}\\
        \big[J^a_{m},J^b_{n}\big] &= \frac{c}{12}m\,\delta^{ab}\delta_{m+n,0} +  i\epsilon^{ab}_{\,\,\,\,c}\,J^c_{m+n} \ ,\label{JJcomm}\\
        \big\{ G^{\alpha}_{\dot{A},r} , G^{\beta}_{\dot{B},s} \big\} &=  \epsilon_{\dot{A}\dot{B}}\bigg[\epsilon^{\alpha\beta}\frac{c}{6}\Big(r^2-\frac14\Big)\delta_{r+s,0} + \big(\sigma^{aT}\big)^{\alpha}_{\gamma}\:\epsilon^{\gamma\beta}(r-s)J^a_{r+s} + \epsilon^{\alpha\beta}L_{r+s} \bigg] \ ,\label{GGcomm}\\
        \big[J^a_{m},G^{\alpha}_{\dot{A},r}\big] &= \h\big(\sigma^{aT}\big)^{\alpha}_{\beta}\, G^{\beta}_{\dot{A},m+r} \ ,\label{JGcomm}\\
        \big[L_{m},G^{\alpha}_{\dot{A},r}\big] &= \Big(\frac{m}{2}  -r\Big)G^{\alpha}_{\dot{A},m+r} \ ,\label{LGcomm}\\
        \big[L_{m},J^a_n\big] &= -nJ^a_{m+n} \ , \label{LJcomm}
    \end{align}
\end{subequations}
where $\sigma^{aT}$ are the transpose of the Pauli sigma matrices and the right-moving modes satisfy an analogous set of relations. A different basis for the $J$ currents is also often used; instead of $J^a$ we have $J^3,J^{\pm}$ where
\begin{equation}
    J^{\pm} = J^1 \pm i J^2 \ .
\end{equation}
We will not have need for the full contracted large $\mathcal{N}=4$ superconformal algebra of the D1-D5 CFT, but this can nonetheless be found in Appendix A of \cite{Guo:2019ady} or Appendix A.4 of \cite{Hampton:2018ygz} with the same conventions. We do, however, give our conventions for the realisation of part of this algebra in terms of the free fermions $\psi^{\alpha A}$ with modes $d^{\alpha A}_r$, and the free bosons $\partial X_{\!A\dot{A}}$ with modes $\alpha_{A\dot{A},n}$. The mode expansions of the fields are given by
\begin{subequations} \label{BosFerModes}
    \begin{align}
        \partial X_{A\dot{A}}(z) &= -i \sum_n z^{-n-1}\,\alpha_{A\dot{A},n} \ ,\\
        \psi^{\alpha A}(z) &= \sum_r z^{-r-1/2}\, d^{\alpha A}_{r} \ ,
    \end{align}
\end{subequations}
and likewise the inverse relations are
\begin{subequations} \label{BosFerModes2}
    \begin{align}
        \alpha_{A\Ad,n} &= i \oint \frac{dz}{2\pi i} z^{n} \partial X_{A\Ad}(z) \ ,\\
        d^{\alpha A}_{s} &= \oint \frac{dz}{2\pi i} z^{n-\frac12} \psi^{\alpha A}(z) \ .
    \end{align}
\end{subequations}
The brackets of the $\alpha$ and $d$ modes are
\begin{subequations}\label{FundComms}
    \begin{align}
        \big[\alpha_{A\dot{A},n},\alpha_{B\dot{B},m}\big] &= - n\frac{c}{6}\epsilon_{AB}\,\epsilon_{\dot{A}\dot{B}}\, \delta_{n+m,0} \ ,\label{eq.alalcomm}\\
        \big\{d^{\alpha A}_{r},d^{\beta B}_{s}\big\} &= - \frac{c}{6}\epsilon^{\alpha\beta}\epsilon^{AB} \delta_{r+s,0} \ ,\label{eq.ddcomm}\\
        \big[\alpha_{A\dot{A},n}, d^{\beta B}_{s}\big] &= 0 \ ,\label{eq.aldcomm}
    \end{align}
\end{subequations}
and likewise for the right-moving fields. The commutators of currents with the $\alpha$ and $d$ modes are
\begin{subequations} \label{eq.currFreeComm}
    \begin{align}
        \big[L_n, \alpha_{A\Ad,m}\big] &= -m \alpha_{A\Ad,m_n} \quad\ ,\qquad \big[L_{n}, d^{\alpha A}_{s}\big] = -\Big(\frac{n}{2}+s\Big) d^{\alpha A}_{n+s} \ ,\label{eq.LFreeComm}\\
        \big[J^a_n, \alpha_{A\Ad,n}\big] &= 0 \quad\ ,\qquad \big[J^a_n, d^{\alpha A}_{s}\big] = \frac12 \big(\sigma^{Ta}\big)^{\alpha}_{\ \beta}\, d^{\beta A}_{s+n} \ ,\label{eq.JFreeComm}\\
        \big[G^{\alpha}_{\Ad,s}, \alpha_{B\Bd,n}\big] &= -in \epsilon_{AB}\epsilon_{\Ad\Bd} d^{\alpha A}_{s+n} \quad\ ,\qquad \big\{G^{\alpha}_{\Ad,r}, d^{\beta B}_{s}\big\} = i \epsilon^{\alpha\beta}\epsilon^{AB} \alpha_{A\Ad,r+s} \ .\label{eq.GFreeComm}
    \end{align}
\end{subequations}
The 2-point functions for the free bosonic and fermionic fields are
\begin{subequations} \label{eq.free2point}
    \begin{align}
        \langle \partial X_{A\Ad}(z_1)\partial X_{B\Bd}(z_2)\rangle &= \frac{\epsilon_{AB}\epsilon_{\Ad\Bd}}{(z_1-z_2)^2} \ ,\label{eq.pdX2point}\\
        \langle \psi^{\alpha A}(z_1) \psi^{\beta B}(z_2) \rangle &= -\frac{\epsilon^{\alpha\beta}\epsilon^{AB}}{z_1-z_2} \ .\label{eq.psi2point}
    \end{align}
\end{subequations}
In terms of these free bosons and fermions the currents are given as
\begin{subequations} \label{eq.currentsFree}
    \begin{align}
        J^a &= \frac14 \epsilon_{\alpha\gamma}\epsilon_{AC}\, \psi^{\gamma C} \big(\sigma^{aT}\big)^{\alpha}_{\ \beta}\psi^{\beta A} \ ,\label{eq.Jfree}\\
        G^{\alpha}_{\Ad} &= \psi^{\alpha A} \partial X_{A\Ad} \ ,\label{eq.Gfree}\\
        T &= \frac14 \epsilon^{AB}\epsilon^{\Ad\Bd} \partial X_{B\Bd}\partial X_{A\Ad} + \frac12 \epsilon_{\alpha\beta}\epsilon_{AB} \psi^{\beta B}\partial\psi^{\alpha A} \ ,\label{eq.Lfree}
    \end{align}
\end{subequations}
with $a=1,2,3$ being an $SU(2)$ vector representation index.

\subsection{Hermitian conjugation} \label{app.Herm}

Suppose we consider an amplitude on the cylinder, in the NS sector, with the form
\be
    A={}_{\scriptscriptstyle{N\!S}}\langle 0|\, {\mathcal O}^\dagger\big( \tau=T, \sigma=0 \big) {\mathcal O}\big( \tau=-T, \sigma=0 \big) |0\rangle_{\scriptscriptstyle{N\!S}} \ .
\label{appone}
\ee
Then we should have $A\ge 0$. This requirement helps determine the way Hermitian conjugates are defined in our CFT. Note that contractions between $su(2)$ indices are done using antisymmetric tensors like $\epsilon_{\alpha\beta}$, and this fact gives rise to certain negative signs in the definitions of Hermitian conjugates. For the supercharges, we use the following rules
\begin{equation} \label{Gconj}
\begin{aligned}
    \Big( G^{+}_{+}(\tau,\sigma)\Big)^{\dagger} &=-G^{-}_{-}(-\tau,\sigma)\quad,\quad \Big( G^{+}_{-}(\tau,\sigma)\Big)^{\dagger}=G^{-}_{+}(-\tau,\sigma) \ ,\\
    \Big( \bar G^{+}_{+}(\tau,\sigma)\Big)^{\dagger} &=-\bar G^{-}_{-}(-\tau,\sigma)\quad, \quad \Big( \bar G^{+}_{-}(\tau,\sigma)\Big)^{\dagger}=\bar G^{-}_{+}(-\tau,\sigma) \ ,
\end{aligned}
\end{equation}
while for the degree-2 twist operators, our conventions are
\be \label{TwistConj}
    \big(\sigma^{--}(\tau,\sigma)\big)^{\dagger}=-\sigma^{++}(-\tau,\sigma)\quad\ ,\qquad \big(\sigma^{-+}(\tau,\sigma)\big)^{\dagger}=\sigma^{+-}(-\tau,\sigma) \ .
\ee
Likewise, in terms of the free fields of the orbifold theory the conjugation conventions we use are
\begin{align} \label{eq.adDagConv}
    \big(\alpha_{++,n}\big)^{\dagger} &= -\alpha_{--,-n} \quad,\quad \big(\alpha_{+-,n}\big)^{\dagger} = \alpha_{-+,-n} \ , \\
    \big(d^{++}_{s}\big)^{\dagger} &= - d^{--}_{-s} \quad,\quad \big(d^{+-}_{s}\big)^{\dagger} =  d^{-+}_{-s} \ .
\end{align}
These conventions ensure that the correlators
\begin{equation}
    \langle \big(\alpha_{A\Ad,n}\big)^{\!\dagger}\, \alpha_{A\Ad,n}\rangle = \langle \big(d^{\alpha A,n}\big)^{\!\dagger}\, d^{\alpha A,n}\rangle =1 \ ,
\end{equation}
for using the brackets \eqref{FundComms}.

\section{Mapping amplitude relations to lift relations} \label{app.B}

In Section \ref{sec.checks} we found interesting relations between amplitudes of certain states and those of their descendants (see \eqref{Lrel} and \eqref{Grel}). Here we describe how to map those amplitude relations to relations between the lifts of states. Consider the following simplified form of the definition of lift given in \eqref{eq.lift}
\begin{align} \label{atoE}
    E^{(2)}(\phi) = \kappa\,\frac{A(\phi)}{\bra{\phi}\ket{\phi}} \ ,
\end{align}
where $\kappa$ is a constant and $A(\phi)$ is an amplitude related to \eqref{eq.Xdef} with initial and final states being $\phi$. Taking, for instance, the amplitude relation 
\begin{equation}
    A(\phi) = 2h_{\psi} A(\psi) \ ,
\end{equation}
obtained in Section~\ref{ssec.Lrelations} by considering the states $|\psi\rangle$ and $\ket{\phi} = L_{-1}\ket{\psi}$ and using \eqref{atoE} we find that at the level of lifts
\begin{equation} \label{eq.LliftRel2}
    E^{(2)}(\phi) = 2h_{\psi}\frac{{\bra{\psi}\ket{\psi}} }{{\bra{\phi}\ket{\phi}}}\, E^{(2)}(\psi) \ .
\end{equation}
Since the norm of the descendant state is given by
\begin{equation}
    \langle\phi|\phi\rangle = \langle\psi|L_1L_{-1}|\psi\rangle = 2\langle\psi|L_0|\psi\rangle = 2h_{\psi} \langle\psi|\psi\rangle \ ,
\end{equation}
we can simplify \eqref{eq.LliftRel2} to give the lifting relation
\begin{equation}\label{eq.LliftRel}
    E^{(2)}(\phi) = E^{(2)}(\psi) \ ,
\end{equation}
\textit{i.e.} that the lift of a state satisfying $L_1|\psi\rangle=0$ is equal to the lift of its $L_{-1}$ descendant. Equally one can start with the amplitude relation
\begin{equation}
    A^{\beta\alpha}_{\Bd\Ad}(\widetilde{\phi}) = K^{\beta\alpha}_{\Bd\Ad} A(\widetilde{\psi}) \ ,
\end{equation}
with $K^{\beta\alpha}_{\Bd\Ad}$ defined in \eqref{Kdef} obtained by considering the states $|\widetilde{\psi}\,\rangle$ and $|\widetilde{\phi}\,\rangle^{\alpha}_{\Ad} = G^{\alpha}_{\Ad,-\frac12}|\widetilde{\psi}\,\rangle$. This can be recast in terms of lifts using \eqref{atoE} and imposing that the index $\beta$ is opposite to $\alpha$ and $\Bd$ opposite to $\Ad$, giving
\begin{equation} \label{eq.GliftRel2}
    E^{(2)}\big(\widetilde{\phi}^{\,\alpha}_{\Ad}\big) = K^{\alpha}_{\!\Ad}\: \frac{\langle\widetilde{\psi}\,|\widetilde{\psi}\,\rangle}{{}^{\,\alpha}_{\Ad}\langle\widetilde{\phi}\,|\widetilde{\phi}\,\rangle^{\alpha}_{\Ad}}\, E^{(2)}\big(\widetilde{\psi}\big) \ ,
\end{equation}
where there is no implied summation over the repeated indices $\alpha$ and $\Ad$ and $K^{\alpha}_{\Ad}$ is defined as $K^{\beta\alpha}_{\Bd\Ad}$ with the above conditions imposed on $\beta$ and $\Bd$. Since the norm of the descendant state is given by (there no sum over repeated indices)
\begin{equation}
    {}^{\,\alpha}_{\Ad}\langle\widetilde{\phi}\,|\widetilde{\phi}\,\rangle^{\alpha}_{\Ad} = \langle\widetilde{\psi}\,|\big(G^{\alpha}_{\Ad,-\frac12}\big)^{\!\dagger}G^{\alpha}_{\Ad,-\frac12}|\widetilde{\psi}\,\rangle = K^{\alpha}_{\Ad}\langle\widetilde{\psi}\,|\widetilde{\psi}\,\rangle \ ,
\end{equation}
we can simplify \eqref{eq.GliftRel2} to give the lifting relation
\begin{equation}\label{eq.GliftRel}
    E^{(2)}(\widetilde{\phi}^{\alpha}_{\Ad}) = E^{(2)}(\widetilde{\psi}) \ ,
\end{equation}
\textit{i.e.} that the lift of a state satisfying $\big(G^{\alpha}_{\Ad,-\frac12}\big)^{\!\dagger}|\widetilde{\psi}\,\rangle=0$ is equal to the lift of its $G^{\alpha}_{\Ad,-\frac12}$ descendant.

Given a chosen basis of states for which the lift is being computed, \textit{i.e.} the set used in this paper
\begin{equation} \label{eq.basisStates}
    \bigg\{\,|\alpha\alpha\rangle_{B\Bd A\Ad(m,n)}\ ,\ \ |\alpha d\rangle^{\alpha A}_{B\Bd(n,s)}\ ,\ \ |dd\rangle^{\beta B\alpha A}_{r,s}\, \bigg\} \ ,
\end{equation}
the descendant states $|\phi\rangle$ and $|\widetilde{\phi}\,\rangle^{\alpha}_{\Ad}$ can generally be written as a sum of basis states. The lift of the descendant state (the left-hand side of the relations \eqref{eq.LliftRel} and \eqref{eq.GliftRel}) can not then be immediately written in terms of the lifts of basis states from \eqref{eq.basisStates}, which are the data points computed in Section~\ref{sec.main}. 

In the case that the descendant state is a sum of two basis states, \textit{i.e.} if we have
\begin{equation} \label{eq.phi12}
    |\phi\rangle = |\phi_1\rangle + |\phi_2\rangle \ ,
\end{equation}
where $|\phi_1\rangle$ and $|\phi_2\rangle$ are states in the set \eqref{eq.basisStates}, then we have
\begin{align} \label{eq.Ephi12}
    E^{(2)}(\phi) &= E^{(2)}(\phi_1+\phi_2) \nonumber\\
    &= \kappa\, \frac{A(\phi_1+\phi_2)}{\langle \phi_1+\phi_2|\phi_1+\phi_2\rangle} \nonumber\\
    &= \kappa\, \frac{A(\phi_1) + A(\phi_2) + 2A(\phi_1;\phi_2)}{\langle\phi_1|\phi_1\rangle + \langle\phi_2|\phi_2\rangle + 2\langle\phi_1|\phi_2\rangle} \nonumber\\
    &= \frac{\langle\phi_1|\phi_1\rangle E^{(2)}(\phi_1) + \langle\phi_2|\phi_2\rangle E^{(2)}(\phi_2) + 2\sqrt{\langle\phi_1|\phi_1\rangle\langle\phi_2|\phi_2\rangle}\, E^{(2)}(\phi_1;\phi_2)}{\langle\phi_1|\phi_1\rangle + \langle\phi_2|\phi_2\rangle + 2\langle\phi_1|\phi_2\rangle} \ .
\end{align}
Here $A(\phi_f;\phi_i)$ is a generalisation of the integrated amplitude $A(\phi)$ used in lifting computations where the initial and final states are not necessarily equal, with $A(\phi_i;\phi_i) = A(\phi_i)$ and $E^{(2)}(\phi_f;\phi_i)$ is the associated lift-like quantity defined analogously to \eqref{atoE} as
\begin{align}\label{atoE2}
    E^{(2)}(\phi_f;\phi_i) \equiv \kappa\,\frac{A(\phi_f;\phi_i)}{\sqrt{\bra{\phi_f}\ket{\phi_f}\bra{\phi_i}\ket{\phi_i}}} \ .
\end{align}
Thus in the case that the descendant state splits as \eqref{eq.phi12} then \eqref{eq.Ephi12} can be used as the left-hand side of the relations \eqref{eq.LliftRel} and \eqref{eq.GliftRel}. This argument can easily be generalised to the case of $|\phi\rangle = \sum_{k=1}^{K} |\phi_k\rangle$ with the states $|\phi_k\rangle$ being in the set \eqref{eq.basisStates}.

\bibliographystyle{JHEP}
\bibliography{twomodelift.bib}
\end{document}